\documentclass[12pt]{article}
\usepackage{graphicx,amsbsy,latexsym,calc,graphics,psfrag}
\usepackage{amssymb}

\def\d{\delta}
\def\e{\epsilon}

\def\r{\rho}

\def\G{\Gamma}

\def\d{\delta}
\def\e{\epsilon}

\def\r{\rho}

\def\G{\Gamma}

\newcommand{\eqref}[1]{(\ref{#1})}
\newcommand{\be}{\begin{equation}}
\newcommand{\ee}{\end{equation}}

\newcommand{\one}{\hbox{l\kern-2mm 1}}

\newcommand{\ba}{\begin{array}}
\newcommand{\ea}{\end{array}}

\def\be{\begin{equation}}
\def\ee{\end{equation}}
\def\bea{\begin{eqnarray}}
\def\eea{\end{eqnarray}}
\newcommand{\bm}[1]{\mbox{\boldmath $#1$}}

\begin{document}
\begin{center}
\Large{Unitarity, Analyticity and Crossing Symmetry  in Two- and Three-hadron Final State Interactions}
\end{center}
\begin{center}
\Large{Ian J. R. Aitchison}
\end{center}
\begin{center}
SLAC National Accelerator Laboratory, Menlo Park, CA 94025, USA

\vspace{2in} 
{\bf Abstract}
\end{center}

These notes are a fuller  version of four lectures given at the 2015 International Summer Workshop in Reaction Theory held at Indiana University, Bloomington. The aim is to provide a simple introduction  to how the  tools of ``the $S$-matrix era'' - i.e. the constraints of unitarity, analyticity and crossing symmetry - can be incorporated into analyses of final state interactions in two- and three-hadron systems. The main focus is on corrections to the isobar model in three-hadron final states, which may be relevant once more as much larger data sets  become available. 

\newpage
\section{Introduction}

These lectures aim to give a simple introduction to the application of unitarity, analyticity and crossing symmetry - the main principles of $S$ - matrix theory (Eden {\em et al.} \cite{ELOP}) - 
to the analysis of final state interactions in two- and three-hadron systems. 

$S$ - matrix theory flourished in the late 1950s and on through the 1960s. It was developed as a theory of the strong interactions between hadrons, to which the perturbative procedures of quantum field theory seemed inapplicable. It is fair to say that $S$ - matrix theory had only limited success as a first principles technique for calculating strong interaction amplitudes, though some important products survive, such as Regge theory (and remarkably enough $S$ - matrix theory gave birth to string theory). Of course, strong interactions came to mean QCD, where both perturbative and non-perturbative (lattice) techniques have been very successful. Nevertheless, {\em ab initio} calculations of few hadron dynamics present a challenge, and $S$-matrix principles remain as valid constraints which should be incorporated into phenomenological analyses.

Although some features, such as isolated resonances, show up clearly on simple intensity plots, in many cases we are interested in more subtle questions related to {\em phases} of amplitudes. Such information will have to come, as usual in quantum mechanics, from interferences. I briefly outline two (oversimplified) examples. 

Suppose we want to study two excited nucleon states ${\rm N}_1^*$ and ${\rm N}_2^*$ which are produced from an initial $\pi {\rm N}$ state, and which decay sequentially to $\pi \pi {\rm N}$ via the two decay chains ${\rm N}_1^* \to \pi \Delta \to \pi \pi {\rm N}$ and ${\rm N}_2^* \to \rho {\rm N} \to \pi \pi {\rm N}$.   Then a simple model 
(essentially the {\em isobar model}) for the amplitude  leading to the final $\pi \pi {\rm N}$ state takes the form 
\be 
F=A_1 t_1 + A_2 t_2 
\label{eq:int1} 
\ee 
where the $A_i$ are the strong production amplitudes for ${\rm N}_1^*$ and ${\rm N}_2^*$, and the $t_i$ are the two-body final state interaction amplitudes in the $\Delta$ and $\rho$ channels. Then $|F|^2$ will contain an interference term proportional to 
$\cos (\theta + \phi)$, where $\theta$ is the relative phase of $A_1$ and $A_2$, and $\phi$ is the relative phase of $t_1$ and $t_2$. So from the intensity $|F|^2$ we can learn about the relative phase of the production amplitudes, {\em provided that} we know the relative phase of the final state two-body  amplitudes $t_1$ and $t_2$. In the isobar model, these are assumed to be determined from the known two-body scattering data.  But we will see that unitarity (or equivalently rescattering amongst the final state particles) forces corrections to the isobar model, which affect these relative phases. At some point, therefore, such corrections should be incorporated into the analysis.    

A second example concerns the extraction of CP-violating phases in states decaying weakly to hadronic final states. Again we can use (\ref{eq:int1}) to make the point, where now 
$A_i$ are the weak production amplitudes and as before the $t_i$ are strong two-body final state amplitudes. The CP-conjugate amplitude will be 
\be 
{\bar F} = A^*_1 t_1 + A^*_2 t_2, 
\ee 
and the CP-violation will be observable from the difference 
\be 
|F|^2 - |{\bar F}|^2 = 4 \, {\rm Im} (A_1 A_2^*) \, {\rm Im} ( t_2 t_1^*). 
\ee 
To get an effect, there needs to be a phase difference between both the two weak amplitudes and the two strong amplitudes. And to extract the value of the CP-violating weak phase difference we need to be sure of the strong phase difference. In two-body final states the latter is known from two-body data, but in three-body states rescattering effects will again modify the $t_i$ phases. 

It would be nice if we could have a phenomenology that was independent of approximations necessarily made in describing the hadronic final state interactions. Such a model-independent analysis generally requires very large data samples. Although these may now be beginning to be available, it seems likely that amplitudes with some theory ingredients will still be needed for some time. And with vastly more data, the deficiencies in models like the isobar model may need to be remedied. A reasonable way to tackle this is to require as a ``minimum theory'' that our amplitudes satisfy the old $S$ - matrix principles mentioned previously - that is, we aim to provide amplitudes which at least obey the constraints of unitarity, analyticity, and crossing symmetry, as far as possible. 

These lectures will describe what these constraints are and how they are implemented in some simple examples. We begin with two-hadron final states, introducing unitarity and the $K$-matrix. Then we add analyticity, and dispersion relations. Our main focus, though, will be on three-hadron final states. We show how unitarity in the two-body sub-energy channels places a constraint on the isobar decay amplitudes, and how analyticity enables us to convert this into  integral equations for modified isobar amplitudes, which satisfy two-body unitarity.  We shall see that, somewhat surprisingly, these amplitudes can actually satisfy three-body unitarity as well. This ``two-body'' approach to what is after all a three-body problem is conceptually very simple, and produces amplitudes which can  directly replace the conventional isobar amplitudes. The price to be paid {\em en route} is a certain amount of gymnastics in the complex plane.

 Throughout we shall restrict ourselves to the simplest possible spin and angular momentum configurations, so that the logic of ``unitarity + analyticity + crossing symmetry'' can be clearly exhibited, unencumbered by other complications. However,   I shall briefly report on the results of calculations from the 1970s and 1980s made for various physically realistic three-hadron systems. But this is not a review: rather, it mostly  describes work that I was myself involved with, and no attempt is made to be comprehensive.

\section{Elastic 2$\to$ 2 Unitarity}
\subsection{One channel, one resonance}
\subsubsection{Unitarity}
The unitarity relation for the $T$-matrix is 
\be
T - T^\dagger= 2 {\rm i} T \rho T^\dagger = 2 {\rm i} T^\dagger \rho T \label{eq:unitT}
\ee
where $\rho$ is the appropriate intermediate state phase space. For simplicity we 
consider the elastic scattering of two identical spinless bosons of unit mass, interacting in the $l=0$ partial wave only. Then (\ref{eq:unitT}) becomes 
\be 
T(s) - T^*(s) = 2 {\rm i} \r (s) |T(s)|^2 \label{eq:unitT2}
\ee
or equivalently
\be{\rm{Im}} T(s) = \r (s) |T(s)|^2 \label{eq:unitT3}
\ee
where $s=4+4q^2$ is the square of the total c.m. energy, $q$ is the c.m. momentum, and (in a convenient normalization) 
\be 
\r (s) =  \left( \frac{s-4}{s} \right) ^{1/2}. \label{eq:rho(s)}
\ee
More generally, for a two-body threshold with unequal masses $m_1$ and $m_2$ the phase space would be  
\be 
\frac{1}{s} {[s-(m_1+m_2)^2][s-(m_1-m_2)^2]}^{1/2} \equiv 
\frac{1}{s} k(s, m_1^2, m_2^2) \label{eq:genrho}
\ee
where 
\be 
k(a, b, c) = a^2 + b^2 + c^2 - 2 ab - 2 bc - 2 ca. \label{eq:defk}
\ee

A parametrisation satisfying (\ref{eq:unitT3}) is 
\be 
T(s) = {\rm e} ^{\rm i \d } \sin \d /  \r = (\r \cot \d - \rm i \r )^{-1} \label{eq:Tdelta}
\ee 
where $\d $ is the phase shift. In particular, if we choose $\d = \tan^{-1} [\r g^2 /(s_{\rm r} -s)] $ the phase shift will rise from zero at threshold to $\pi$ as 
$s \to \infty$, passing through $\pi /2$ at $s=s_{\rm r}$. This is a standard Breit-Wigner type resonance formula, with amplitude 
\be 
f(s) = \frac{g^2}{s_{\rm r} - s - \rm i \r (s) g^2}.\label{eq:BW} 
\ee
Near the peak of a narrow resonance we may set $\r (s) \approx \r (s_{\rm r})$;   then the resonance maximum is reached at $s=s_{\rm r}$, and the full width at half height in a  plot of $|T(s)|^2$ versus $s$ is $2 \r (s_{\rm r}) g^2 $. We represent $f(s)$ by figure 1.

\begin{figure}
\begin{center}
\includegraphics{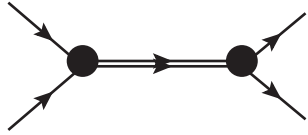}
\end{center}
\caption{Resonance amplitude}
\end{figure}

\subsubsection{The complex plane, Reimann sheets}

We shall frequently be considering variables such as $s$ to be complex, and our amplitudes will be assumed to be {\em analytic functions} of their arguments\footnote{Burkhardt's book \cite{hugh} contains a useful long first section on complex variable analysis, and then continues with equally relevant sections on collision theory and $S$-matrix dynamics.}. An example immediately arises in the case of the function $\rho(s)$ in (\ref{eq:unitT}). In the unitarity relation as written in (\ref{eq:unitT}) it is implicit that $s \geq 4$, the elastic scattering threshold. In that case, $\rho(s)$  should be multiplied by $\theta(s-4)$. But this is not an analytic function of $s$. Rather, we shall understand (\ref{eq:unitT}) to be true as it stands, and allow $\rho(s)$ to be defined for  all values in the complex $s$ plane, by analytic continuation from the physical region. That region is the real axis $s \geq 4$, approached from above: $\lim_{\epsilon \to 0}f(s+\rm i \epsilon)$. We need to be careful how we approach the real axis because, as we now discuss, it makes a difference due to the singularity structure of $\rho(s)$.

 Viewed as an analytic function of the complex variable $s$, $\rho(s)$ has branch points at $s=0$ and $s=4$, with associated cuts as shown in figure 2. 
 \begin{figure}
 \begin{center}
 \includegraphics{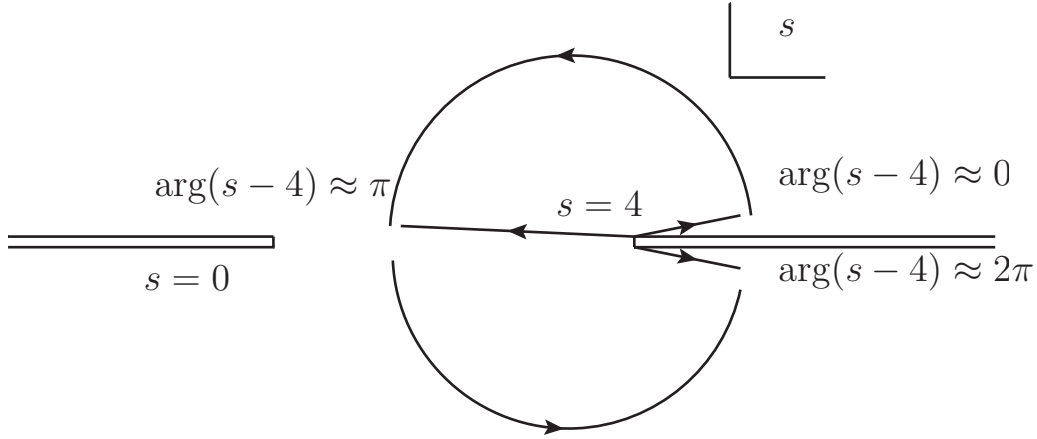}
 \end{center}
 \caption{The branch points and cuts of $\rho(s)$}
 \end{figure}
 The physical region for our 2-particle scattering is the real axis $s \geq 4$. Suppose we start at a point just above the real axis, with $s > 4$, with the square root function defined to be positive.    We can continue the function $\sqrt{s-4}$ on a circular path encircling the point $s=4$, starting at a point just above  the real axis $s > 4$, passing between $s=4$ and $s=0$, and returning to the real axis to the right of $s=4$ but just below the real axis. On the real axis in the region $0 < s < 4$ the square root becomes $\rm i \sqrt{4-s}$, and at the end of the trip it has become $-\sqrt{s-4}$. Notice that the value of the function $\sqrt{s-4}$ for $s>4$ and just above the real axis is not the same as the value of $\sqrt{s-4}$ for $s>4$ and just below the real axis. That is why we draw a ``cut'' along the real axis $s \geq 4$, to remind ourselves of this {\em discontinuity} in the function $\sqrt{(s-4)}$. It is the value reached from above the real axis that is the ``physical limit''.  

In addition to the branch point at $s=4$, $\rho(s)$ also has a branch point at  at $s=0$. Whereas the branch point at $s=4$ has a clear physical origin - namely the two-particle threshold - that at $s=0$ does not: it may be called a ``kinematic'' singularity. We shall see in section 4.1 how to get rid of it.
We therefore continue to focus on the square root branch point at $s=4$. 

We have discussed making one complete circuit of the point $s=4$, ending up just below the real axis, to the right of $s=4$. Let's continue on from this point, and make another complete circuit. On this second circuit, the argument of $s-4$ starts at the value $2\pi$, and ends at the value $4\pi$. Half-way round 
this second circuit, the argument has the value $3\pi$. For the square root, we have to halve the argument, so the square root function starts at the value $-\sqrt{s-4}$, becomes $-{\rm i} \sqrt{4-s}$ 
half way round, and ends at the value $\sqrt{s-4}$ after the complete (second) circuit.  Thus after two complete circuits around $s=4$ the square root function returns to its original value. 

This description has been in terms of a double-valued function $\pm \sqrt{s-4}$ defined over a single complex plane. The standard alternative description shifts the multi-valuedness from the function to the space over which it is defined. In the present case, we will  have {\em two} complex planes, called ``sheets''. On the first sheet, we use the positive square root $+\sqrt{s-4}$, and on the second sheet we use the negative square root $-\sqrt{s-4}$. On each sheet, we are dealing with a single-valued function. 
The interesting thing is that the sheets are connected in the region of the cut. Going once around $s=4$ on sheet I, say, the function $\sqrt{s-4}$ ends up at the value $-\sqrt{s-4}$, which is just the same as the value of the function on the second sheet (namely $-\sqrt{s-4}$) evaluated just above the cut. So after one revolution on sheet I we pass smoothly onto sheet II as we cross the real axis. Continuing round on sheet II, we arrive after one circuit at a point just below the real axis $s>4$, where the second sheet function takes the value $+\sqrt{s-4}$. This is the same as the value we started with in sheet I, before the two circuits. So the second time we cross the axis to the right of $s=4$, we are back on sheet I. The way the sheets are connected along the cut is indicated in figure 3. 

\begin{figure} 
\begin{center} 
\includegraphics[scale=.7]{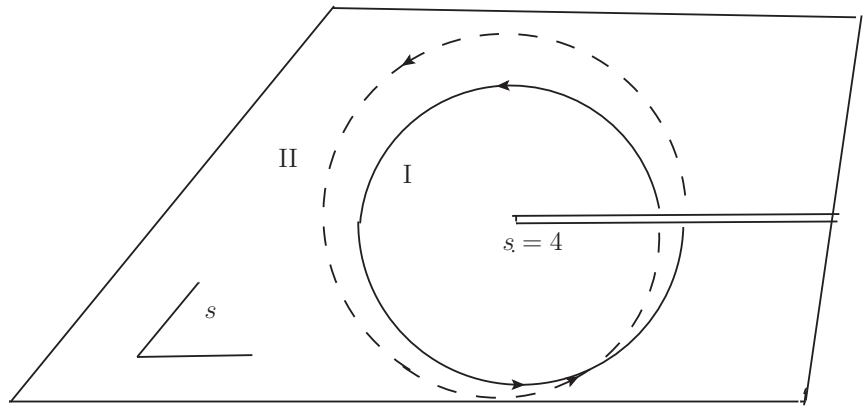}
\end{center}
\caption{The two-sheeted function $\sqrt{s-4}$: a circuit on sheet I followed by a circuit on sheet II 
gets back to sheet I.}
\end{figure}

Another description of the square root function is also possible, and perhaps easier to visualize. Since the ``sheet'' business is all to do with the square root, maybe things would be simpler if we introduced a new variable which is the square root itself, rather than $s$: namely, define a new variable 
\be 
q=\frac{1}{2} \sqrt{s-4}, \label{eq:q}
\ee
which is of course just the magnitude of the momentum. If we set $s=4+r{\rm e}^{{\rm i}\theta}$, which parametrises a circle with centre at $s=4$ and radius $r$, then $q=\frac{1}{2} r^{1/2} {\rm e}^{{\rm i} \theta/2}$. It follows that the whole of the {\em first} $s$-sheet with $0 \leq \theta \leq 2\pi$ corresponds to the {\em upper} half $q$-plane, while the whole of the {\em second} $s$-sheet ($2 \pi \leq \theta \leq 4\pi$) corresponds to the {\em lower} half $q$-plane. Two trips around the $s=4$ threshold in $s$ correspond to just one trip around $q=0$, as shown in figure 4. 
\begin{figure} 
\begin{center} 
\includegraphics{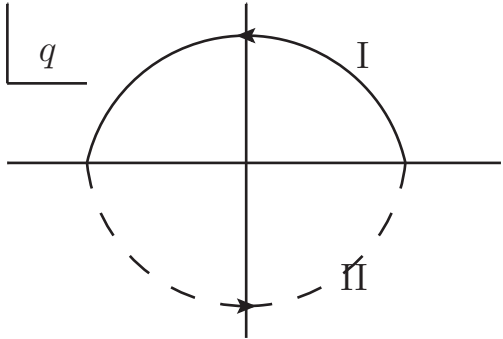}
\end{center}
\caption{The upper half-circuit I in the $q$-plane is equivalent to the first full circuit I in the $s$ plane, and the lower half-circuit II in $q$ is equivalent to the  second circuit II in $s$. }
\end{figure} 

\subsubsection{Resonance poles} 

There is nothing more to be said about the square root function. What about the singularities of $f(s)$ as given by (\ref{eq:BW})? It will simplify matters if  the branch point at $s=0$ is not present. We will see how to get rid of it in section 4.1, but for the moment it will be  sufficient to replace the phase space factor $2q/\sqrt s$ by $2q/\sqrt s_{\rm r}$.  So we  consider the amplitude 
\be 
f_{\rm R} (s) = \frac{\sqrt s_{\rm r} \gamma}{q_0^2 + \gamma^2 -q^2 - 2 {\rm i} \gamma q} = \frac{\sqrt s_{\rm r}}{2q_0} \left \{ 
\frac{\gamma}{q+q^*_{\rm R}} - \frac{\gamma} { q-q_{\rm R}} \right \} 
\label{eq:BWq}
\ee
where $s_{\rm r} = 4 + 4q_0^2 + 4 \gamma^2$ and 
\be
q_{\rm R} = q_0 -{\rm i} \gamma. \label{eq:qR}
\ee 
This $f_{\rm R}(s)$ is essentially the same as $f(s)$ of (\ref{eq:BW}), but without the $\sqrt{s}$ singularity in $\rho(s)$. It satisfies the unitarity relation (\ref{eq:unitT3}) with the phase space factor $2q/ \sqrt{s_{\rm r}}$. In addition to the branch point at $s=4$, $f_{\rm R}(s)$ has poles at $q=q_{\rm R}$ and $q=-q_{\rm R}^*$, as shown in figure 5, both of which have negative imaginary parts. It follows that such a resonant amplitude has two poles in the {\em second} $s$-sheet. Bearing in mind that the physical region is just above the real $s \geq 4$ axis in the first $s$-sheet, which is also just above the real $q \geq 0$ axis in $q$, we see that the pole at $q_{\rm R}$ in the second $s$-sheet is {\em near} the physical region, but the pole at $-q_{\rm R}^*$ on the second $s$-sheet is {\em far} from the physical region (in the sense of distance travelled in the complex plane). These poles in $s$ are at $s=4+4q_{\rm R}^2$ (position A in figure 5) and at $s=4+q_{\rm R}^{*2}$ (position B in figure 5). 
\begin{figure} 
\begin{center} 
\includegraphics{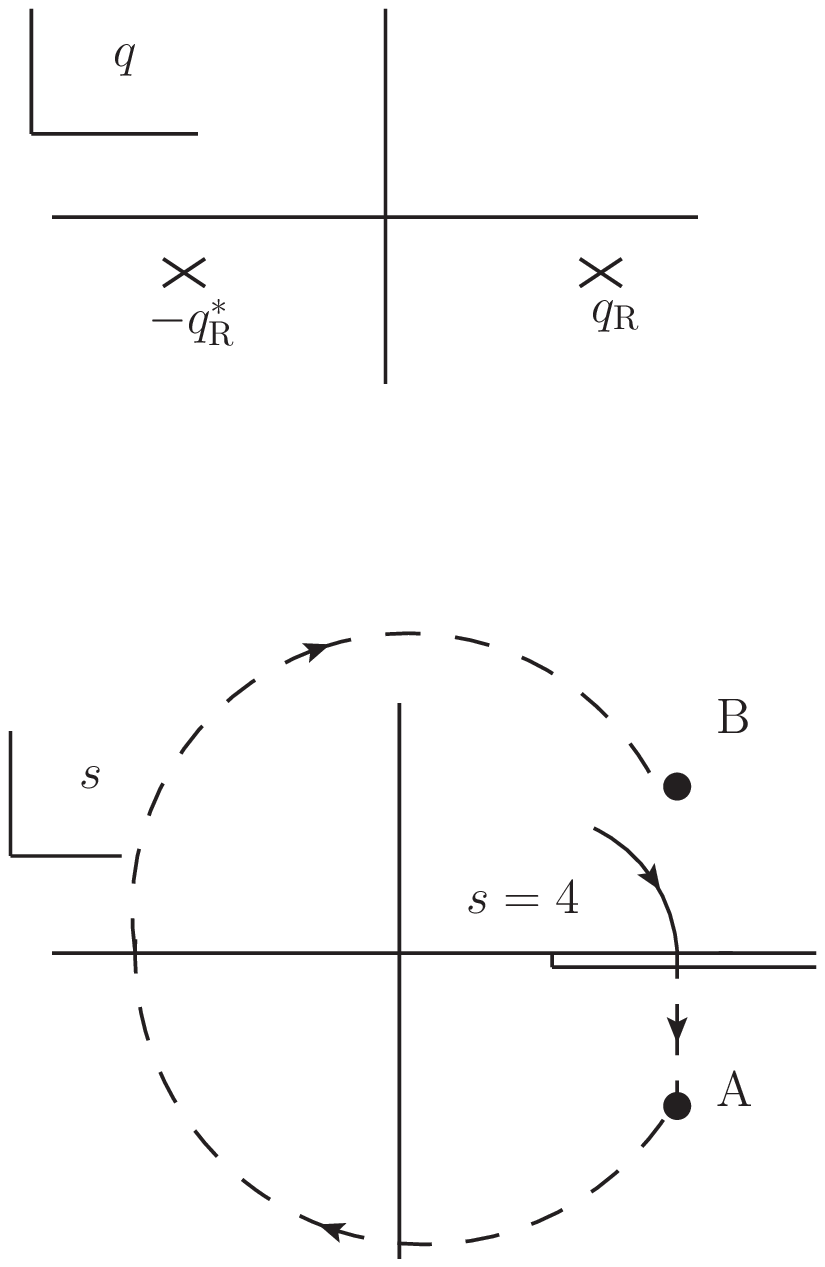}
\end{center}
\caption{The resonance poles in the lower half $q$-plane both correspond to poles in sheet II in the $s$-plane; only the pole at position A is close to the physical region. }
\end{figure}

\subsubsection{Unitarity and discontinuities}

The functions $f(s)$ and $f_{\rm R}(s)$ (and all amplitudes we deal with) satisfy an important condition called {\em Hermitian analyticity}: 
\be 
f^*(s)=f(s^*).\label{eq:HA}
\ee 
Consider for example the function $f_{\rm R}(s)$, and take a point just above the cut at position $s+{\rm i} \epsilon$. Here the square root function (on the first sheet) takes the value $+\sqrt{s-4}$ as $\epsilon \to 0$, and 
\be 
f_{\rm R}(s+{\rm i} \epsilon) = \frac{ \gamma \sqrt{s_{\rm r}}} { q_0^2 + \gamma^2 - q^2 - 2 {\rm i} q}. \label{eq:fR+}
\ee 
On the other hand, $f_{\rm R}$ evaluated at the complex conjugate position $s-{\rm i} \epsilon$ is 
\be 
f_{\rm R}(s - {\rm i} \epsilon) = \frac{ \gamma \sqrt{s_{\rm r}}}{q_0^2 + \gamma^2 - q^2 + 2 {\rm i} \gamma q},  \label{eq:fR-} 
\ee 
because, as we saw, the square root function takes the value $- \sqrt{s-4}$ at a point just under the cut. So clearly 
\be 
f_{\rm R}(s -{\rm i} \epsilon) = f^*_{\rm R}(s+{\rm i} \epsilon) \label{eq:Herman} 
\ee 
and (\ref{eq:HA}) is satisfied. 

Applied to the amplitude $T(s)$ of section 2.1.1, the Hermitian analyticity condition allows us to rewrite the unitarity condition (\ref{eq:unitT2}) as 
\be 
T_+ - T_- = 2 {\rm i} \r T_+ T_- \label{eq:discT}
\ee
where $T_{\pm} = T(s \pm \rm i \e)$, and $\r$ on the RHS of (\ref{eq:discT}) is understood to be $\r_+$. The LHS of (\ref{eq:discT}) is the difference between the values of $T$ just above and just below the $s \geq 4 $ cut: it is the {\em discontinuity} of $T$ across the cut. Rewriting unitarity equations as discontinuity relations will be an essential tool when we come to combine unitarity with analyticity by writing dispersion relations for our amplitudes. We represent (\ref{eq:discT}) diagrammatically by figure 6, where the lines with dots on are ``on-shell'' - i.e. they are physical intermediate state particles, not Feynman propagators.  

\begin{figure}
\begin{center}
\includegraphics[scale=.8]{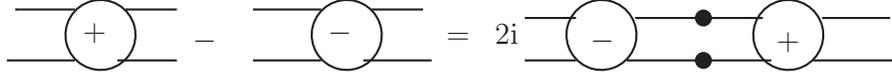}
\end{center}
\caption{The discontinuity relation (\ref{eq:discT}).}
\end{figure}

\subsubsection{The ${ K}$-matrix} 

Dividing both sides of (\ref{eq:discT}) by $T_+ T_-$ we find 
\be 
T_+^{-1} - T_-^{-1} = - 2 \rm i \r \label{eq:discT-1}
\ee
which is another  way of writing the unitarity condition. Since, as we have seen, 
\be 
\r_+ - \r_- = 2 \r   \label{eq:discr}
\ee 
(\ref{eq:discT-1}) may be satisfied by simply writing 
\be 
T^{-1} = K^{-1} - \rm i \r \label{eq:T-1K-1}
\ee 
where $K^{-1}$ has no branch point (is a regular function) at $s=4$. Comparing (\ref{eq:T-1K-1}) with 
(\ref{eq:Tdelta}) we can identify $K^{-1}$ with $\r \cot \d $. And if we choose 
\be 
K=g^2 / (s_{\rm R} - s ) \label{eq:Kres}
\ee 
we recover the B-W amplitude (\ref{eq:BW}). Equally, we can take $K=  \gamma \sqrt{s_{\rm r}}/(q_0^2 + \gamma^2 - q^2)$ with $\rho =2q/ \sqrt{s_{\rm r}}$ and recover $f_{\rm R}(s)$. In general, 
\be 
T=K(1-{\rm i} \r K)^{-1}  = (1- {\rm i} K \r )^{-1} K. \label{eq:TK} 
\ee 
$T$ will satisfy (\ref{eq:unitT}) if $K$ is real. In the case of a resonance, we may think of $K$ as representing a bound state, coupling to the initial and final states with coupling $g$, the factor $(1- \rm i \r K)^{-1}$ then accounting for  the state's decay to the open 2-body channel. 

It is important to note that while (\ref{eq:discr}) is certainly true, the function $\r (s)$ is by no means the only one that has the required discontinuity $2 \r $. In section 4.1 we will see how to manufacture a function that has this discontinuity but does not have the kinematical singularity at $s=0$. 

\subsection{Several channels and resonances} 

Suppose now that we have two resonances, but still only one channel. We might think of adding two B-Ws together to form the amplitude 
\be 
f(s) = \frac{g_1^2}{s_1-s-{\rm i} \r g_1^2} + \frac{g_2^2}{s_2-s-{\rm i} \r g_2^2}. \label{eq:2BW} 
\ee 
But you can soon convince yourself that this will not satisfy the unitarity constraint (\ref{eq:unitT}).
A little more work shows that the violations of (\ref{eq:unitT}) are of order $|g^2/(s_1 - s_2)|$, where $g$ is the larger of $g_1$ and $g_2$. So if the resonances are narrow and well separated, adding the two B-Ws will be a reasonable approximation. In cases where the resonances have more overlap, we can ensure unitarity by putting the two states into $K$ and letting the machinery see to unitarity:
\be 
K= \frac{g_1^2}{s_1-s} + \frac{g_2^2}{s_2 - s}; \label{eq:2poleK}
\ee 
inserting this into (\ref{eq:TK}) guarantees a unitary $T$. 

This formalism really shows its usefulness when more than one channel is open. The quantities $K$ and $T$ now become matrices in the space of channels, and so does $\r $ which is a diagonal matrix of the form 
\be 
\left( \ba{ccc} \r_1 &0& 0 \\
0 &\r_2& 0 \\
0 &0&  \r_3 
\ea \right) 
\ee  \label{eq:rhodiag}
in a 3-channel case, for example, with $\r_1 \sim [s-(m_1 + m_2)^2]^{1/2}$ and $ \r_2 \sim [s-(m_3 + m_4)^2]^{1/2}$, etc. Equation (\ref{eq:unitT}) still holds, with $T$ now a matrix , as do 
(\ref{eq:T-1K-1}) and (\ref{eq:TK}). The matrix $K$ is real, and it can be shown that time reversal invariance requires it to be symmetric \cite{dick}. 

For example, in the case of a 1-resonance 2-channel problem, we would set 
\be 
K_{ij} = \frac{g_i g_j}{s_{\rm R} - s} \ \ \ \ \ \ \ (i,j = 1,2) \label{eq:2res1ch}
\ee 
where the $g_i$ represents the coupling of the resonance to channel $i$. Then we find, for example, 
\be 
T_{12} = \frac{g_1 g_2}{ s_{\rm R} -s - {\rm i} g_1^2 \r_1 - {\rm i} g_2^2 \r_2} \label{eq:flatte}
\ee 
which is often called the Flatte form \cite{flatte}. Notice that each of the thresholds in $\r_1$ and $\rho_2$ generates two Reimann sheets, so such an amplitude takes values on four Reimann sheets. 

We can equally easily deal with the case of more than one resonance in more than one channel. For two resonances, we simply set 
\be 
K_{ij} = \frac{g_{ia}g_{aj}}{s_a -s} + \frac{g_{ib}g_{bj}}{s_b-s} \label{eq:2res2ch}
\ee 
and crank the $T=(1-{\rm i} K \rho )^{-1} K$ handle.  

Thus far we have chosen $K$ to represent only one or more resonances. There is nothing to stop us including a non-resonant ``background'' term in $K$, which can be any real function of $s$ without the unitarity-induced branch points. 

\section{Unitarity in Two-Hadron Final State Interactions} 

We now consider an amplitude $F(s)$ represented by figure 7, where there is for the moment only one
\begin{figure}
\begin{center}
\includegraphics{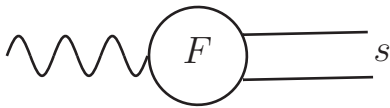}
\end{center}
\caption{Final state interaction amplitude $F(s)$.}
\end{figure}
final state channel, and where the wiggly line could stand for a one-particle state, or for the partial wave projection of a two-particle state amplitude, or   for the projection of a more complicated production amplitude  such as the one shown in figure 8, in which we are going to parametrise
\begin{figure} 
\begin{center}
\includegraphics{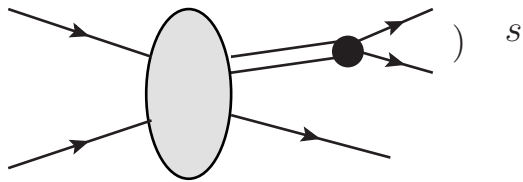}
\end{center} 
\caption{Production process leading to final $s$-channel.}
\end{figure} 
 the blob as just some ``production vertex''. 
 
 Consider for example a decay of the form ${\rm K} \to \pi + \pi$, where the charge labels are irrelevant for the present purpose. We can picture this as proceeding via an initial weak transition to a two-pion state, followed by {\em recattering} of the pions in the final state, because $m_{\rm K}$ is above the two-pion continuum threshold at $2 m_\pi$. In this case $F(s)$ will be the full decay including the 
 strong rescatterings, and $T(s)$ will be the elastic two-body $\pi \pi \to \pi \pi$ amplitude describing the rescatterings. $F(s)$ will here be evaluated at the discrete point $s=m^2_{\rm K}$, but in a process such as that in figure 6, the final state variable $s$ will run continuously over a phase space interval.

 Let's suppose that the two final state particles scatter via a strong interaction $T$-matrix $T(s)$. Then just as in (\ref{eq:discT}) and figure 6, $F(s)$ will satisfy a discontinuity relation 
\be 
F_+ - F_- = 2 {\rm i} T_+ \rho F_- = 2 {\rm i} T_- \rho F_+ \label{eq:discF}
\ee 
which is equivalent to the unitarity constraint 
\be 
{\rm Im} F = T \rho F^* = T^* \rho F. \label{eq:ImF}
\ee 
Relation (\ref{eq:discF}) is represented by figure 9. Since ${\rm Im } F$ must be real, (\ref{eq:ImF})
\begin{figure}
\begin{center}
\includegraphics[scale=.7]{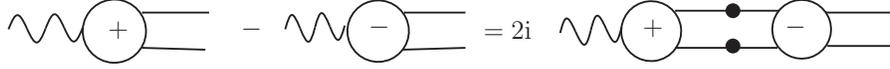}
\end{center}
\caption{Discontinuity relation for $F(s)$.}
\end{figure}
 shows that $F$ must have the phase of $T$. This is an important result, known as Watson's theorem 
\cite{watson}. 

We might suspect that there is a ``$K$-matrix'' type of solution to the unitarity constraint on $F$. Indeed, writing (\ref{eq:discF}) 
\be 
(1-2{\rm i} T_- \rho) F_+ =F_- \label{eq:discf2}
\ee 
and substituting for $T_-$ from (\ref{eq:TK}) we find 
\be 
\frac{1}{1+ {\rm i} K \rho} (1- {\rm i} K \rho) F_+ = F_-,   \label{eq:F+F-}
\ee
which by inspection is satisfied by 
\be 
F_+ = \frac{1}{1-{\rm i} K \rho} P    \label{eq:FP} 
\ee
where $P$ has no branch point at $\rho =0$. 

How would we describe the  production of a single resonance in this formalism? We know that for $K$ we would take $K=g^2/(s_{\rm R} - s)$. If we were to choose just a constant, say, for $P$, we would end up with a {\em zero} in $F$ at the (supposed) peak position $s_{\rm R}$. Instead, we should take 
\be 
P=\frac{f_p g } {s_{\rm R}-s}   \label{eq:Pres}
\ee 
where $f_p$ represents the coupling of the resonance to the initial state, and $g$ is as before its coupling to the final 2-particle state. A verification that (\ref{eq:Pres}) is the right prescription can be provided via potential theory \cite{me72}. Then of course we obtain
\be 
F=\frac{f_p g}{s_{\rm R} -s - {\rm i} \rho g^2}. \label{eq:Fres}
\ee

The generalisation to the multi-channel case is straightforward: $K$ remains the same matrix as in the elastic multichannel $2 \to 2$ case, and $P$ becomes a column vector with a single channel index, since it describes production from a fixed initial state $p$ to a variety of final states $i$. Thus for the production of several resonances $a = 1, 2, \ldots$ decaying to several final states $i$ we take 
\be 
P_{ip} = \Sigma_a g_{i a} \frac{1}{s_a - s} f_{ap}  \label{eq:Presa}
\ee 
and $F= (1 - {\rm i} K \rho) ^{-1} P$ as a matrix equation. 

All this is quite simple - but, less obviously, we can include background terms in both $K$ and $P$ and still be sure that the end result obeys unitarity. Suppose for example we take 
\be P=\frac{f_p g}{s_{\rm R}-s} + B, \ \ \ \ K= \frac{g^2}{s_{\rm R} -s}    \label{eq: Pback} 
\ee 
in a one-channel problem (with an obvious multi-channel extension). Then 
\be 
F=\frac{f_p g} {s_{\rm R} - s - {\rm i} g^2 \rho} + 
\frac{B(s_{\rm R} -s)}{s_{\rm R} -s - {\rm i} g^2 \rho } = (f_p / g) {\rm e}^{{\rm i} \d } \sin \d + 
B {\rm e}^{{\rm i} \d } \cos \d .    \label{eq:Fres+back}
\ee 
If $B$ is real, the phase of $F$ is still that of the elastic scattering amplitude, but there is now a ``$\cos \d $'' piece, as well as a piece proportional to the BW amplitude. However there is no requirement that B actually has to be real: it might, for instance, be taken  
to be a Deck amplitude \cite{deck}, which involves a complex Reggeised pion exchange production process. This kind of model was used \cite{bowler1} in an early analysis of diffractive production of the $a_1$ (one pole, one channel), the $K_1(1270)$ and $K_2(1400)$ (two poles, two channels), and some $N^*$ states (one pole, two channels). A similar but more elaborate two channel analysis of the $a_1$ was done by Basdevant and Berger \cite{basber}. 

One can also add a term of the form $\alpha B K $ to $P$, which would represent production by the process $B$ followed by elastic scattering via $T$ - i.e. a term 
$\alpha B \exp{{\rm i} \delta} \sin \delta$ in (\ref{eq:Fres+back}). This was done in the analysis of the ACCMOR data by Daum {\em et al.} \cite{daum1}; they concluded that the behaviour of the $\rho \pi$ $L=0$ 
$J^P=1^+$ amplitude could be explained in detail by such a model, in which the Deck amplitude is rescattered through the $a_1$, which may also be directly produced. 

A particularly good (and more recent) example of $K$-matrix methods is the $(IJ)^{PC} = (00)^{++}$ state of $\pi^+ \pi^-$. Anisovich and Sarantsev \cite{ansar} made a fit to the available scattering date from $\pi \pi$ threshold up to $\sqrt{s} = 1.9 \, {\rm GeV}$. They included 5 channels ($i= \pi \pi, K {\bar{K}}, 4 \pi, \eta \eta$ and $\eta \eta'$) and 5 poles in their $K$-matrix. They also added a slowly varying term, and an Adler zero term. The FOCUS collaboration \cite{focus} used this $K$-matrix and formula (\ref{eq:FP}) to describe this $\pi \pi$ wave in a Dalitz plot analysis of $D$ decays to $\pi^+ \pi^- \pi^+$ (only the final $i= \pi \pi$ channel is required from $F$). The $P$ vector contained the same 5 poles, with 5 new coupling parameters $f_{ap}$, and a new background term (but no Adler zero). The collaboration found, in particular, that the low mass $\pi^+ \pi^-$ structure of the $D^+$ Dalitz plot was well reproduced in this $K$-matrix model, without the need of an {\em ad hoc} ``$\sigma$'' term. This shows that the same $K$-matrix description gives a coherent picture of both two-body scattering experiments involving light quark constituents and (heavier quark) charm meson decays. 

A  second interesting application of this  formalism is to the $K \pi \pi$ system in $B$ decays to $K_1(1270) \pi$ and $K_2(1400) \pi$, where the $K$s have $J^P=1^+$ and decay to $K \pi \pi$ \cite{aubert1}. Following the analysis by the ACCMOR collaboration \cite{daum2}, the $ K \pi \pi$ system is described by a $K$-matrix model comprising six channels ($K^*(892) \pi, \rho K, K_0^*(1430) \pi, f_0(1370) K, (K^*(892)\pi)_{L=2}, 
\omega K)$, and two resonances. (We note in parenthesis that these channels (and the ealier $\pi \rho$ one) are only ``quasi two-body'' channels, since one of the two particles is strongly unstable; we shall discuss this point further in section 8). The $K$-matrix parameters were determined from an {\em ab initio} fit to the ACCMOR data from the WA3 experiment ($K^- p \to K^- \pi^+ \pi^- p$ at 
63 GeV), taking 
\be 
P = (1 + \tau K ) B + R \label{eq:babarP}
\ee 
where $R$ is the two-pole direct production term having the form (\ref{eq:2res2ch}), and $B$ is a background term (as usual (\ref{eq:babarP}) is understood to be a vector in channel space). This will generate an amplitude $F$ via (\ref{eq:FP}). In applying the model to $B$ decays, the same $K$-matrix was retained (as in the FOCUS work), but the background $B$ was set to zero, since no strong diffractive production process was now present. 

So far we have been considering parametrisations of the data that respected the constraints of two-body unitarity. We turn now to the inclusion of another ingredient - analyticity. 

\section{Combining Unitarity and Analyticity} 

\subsection{Elastic two hadron  $\to$ two hadron  reactions}

\subsubsection{The Chew-Mandelstam phase space factor}

Let's begin by briefly recalling some simple formulae. Suppose $f(z)$ is analytic in and on a closed contour $C'$ (see figure 10). Then Cauchy's theorem implies that $f(z)$ can be written as 
\begin{figure}
\begin{center} 
\includegraphics{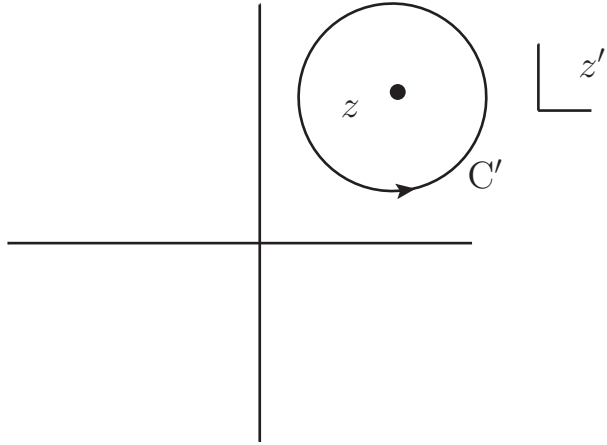}
\end{center} 
\caption{Cauchy's theorem contour.}
\end{figure}
\be 
f(z) = \frac{1}{2 \pi {\rm i}} \int_{C'} \frac{f(z')}{z' - z} {\rm d} z'.   \label{eq:cauchy}
\ee 
Now suppose that $f(z)$ has only one branch point at the real value $s=4$, with a cut attached running along the real axis, $s \geq 4$. Then we can freely distort $C'$, without running into any singularity of $f(z')$, into the contour $C$ shown in figure 11. The representation (\ref{eq:cauchy}) now becomes
\begin{figure}
\begin{center}
\includegraphics[scale=.7]{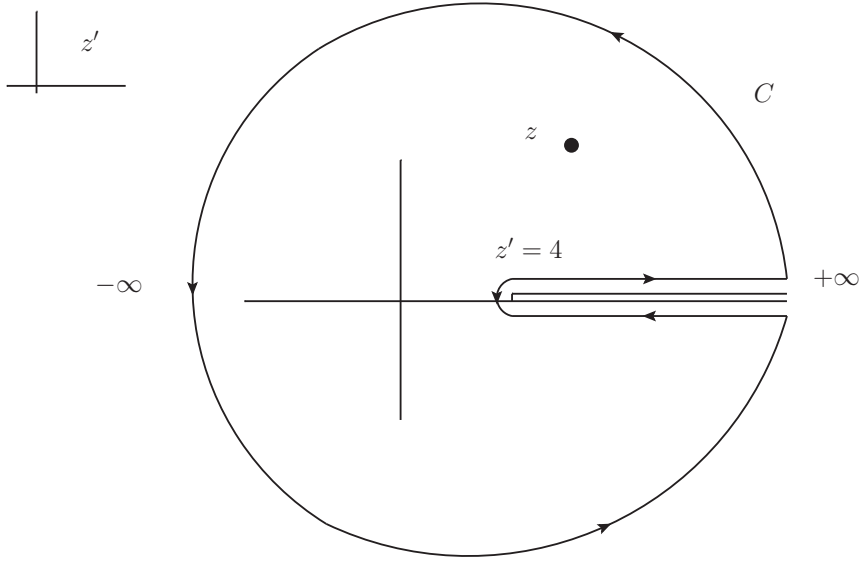} 
\end{center} 
\caption{The distorted contour C, consisting of a circle at infinity and a piece from $+ \infty$ to 4 going below the cut, plus a piece from 4 to $+ \infty$ going above the cut.}
\end{figure} 
\be 
f(z) = \frac{1}{2 \pi {\rm i}} \int_4^\infty \frac{f(s' + {\rm i} \epsilon) - f( s' - {\rm i} \epsilon)}{s' -z}   {\rm d}  s',     \label{eq:fdr}
\ee 
{\em assuming} that convergence is such that we can throw away the part of $C$ at infinity. The numerator of the integrand in (\ref{eq:fdr}) is precisely what we have been calling $f_+ - f_-$, the discontinuity across the $s \geq 4$ cut, which is determined by unitarity. So perhaps we can construct a more complete parametrisation by combining unitarity with analyticity. 

In fact, equation (\ref{eq:discT-1}) tells us that the discontinuity of the inverse of the elastic amplitude is determined only by the phase space factor. Let us see where this leads us. Applying (\ref{eq:fdr}) to $T^{-1}(s)$ we obtain 
\be 
T^{-1}(s) \stackrel{?}{=} - \frac{1}{\pi} \int_4^\infty  \sqrt{{\frac{s'-4}{s'}}} \frac{{\rm d} s'}
{s' -s} \equiv I(s).  \label{eq:T-1dr}
\ee 
Unfortunately the integral diverges logarithmically, but if we are content to input an arbitrary constant into the calculation in the form of the value of $T^{-1}$ at $s=s_0$ we can write 
\bea 
T^{-1}(s) &=& I(s_0) + ( I(s ) - I(s_0)) \\
&=& I(s_0) - \frac{(s-s_0)}{\pi} \int_4^\infty  \sqrt{\frac{s' - 4}{ s'}} \frac { {\rm d} s'}{(s' -s_0)(s'-s)}, 
\eea
where the integral now  converges. It is convenient to take $s_0=4$. Then  we find 
\be
T^{-1} (s) = {\rm constant} + L(s) \label{eq:TL}
\ee
where 
\be 
L(s) = \frac{1}{ \pi} \sqrt{\frac{s-4}{s}} \ln \left( \frac{ \sqrt{s-4} + \sqrt{s}}{-\sqrt{s-4} + \sqrt{s} } \right), 
\ee 
and the logarithm is defined so that its imaginary part is $- \pi$ for $s$ real and greater than 4. A careful study of $L(s)$ shows that despite appearances it does not have the branch point at $s=0$ present in $\rho(s)$. The imaginary part of the logarithm correctly reproduces the unitarity requirement, and the integration in $I(s)$ has banished the singularity at $s=0$ to an unphysical sheet. Functions such as $L(s)$ were introduced by Chew and Mandelstam \cite{chewman}. Note that more generally we would still satisfy unitarity if we replaced ``constant'' in (\ref{eq:TL}) by a regular function $r(s)$ where $r$ has no RH cut and can be identified with $K^{-1}$. This produces a $T(s)$ whose only branch point is at $s=4$, but it may of course have resonance poles on the second sheet reached through the $s \geq 4$ cut. 

\subsubsection{Reconstructing the resonance amplitude $f_{\rm R}(s)$}

A somewhat more complicated exercise in the use of (\ref{eq:fdr}) is provided by the resonance amplitude 
$f_{\rm R}(s)$ of (\ref{eq:BWq}). We have 
\be 
f_{\rm R}(s+{\rm i} \epsilon) = (\gamma \sqrt{s_{\rm r}}/2 q_0) \left \{ \frac{1}{q+q_{\rm R}^*} - \frac{1}{q-q_{\rm R}}\right \}, \ \ \ q_{\rm R} = q_0 - {\rm i} \gamma. \label{eq:fR}
\ee
Along the lower side of the cut, $q$ is replaced by $-q$ so that the discontinuity of $f_{\rm R}$ is 
\bea 
f_{{\rm R}+} - f_{{\rm R}-} &=& (\gamma \sqrt{s_{\rm r}}/2 q_0) \left \{ \frac{1}{q+q_{\rm R}^*} - \frac{1}{q-q_{\rm R}} - \frac{1}{-q + q_{\rm R}^*} + \frac{1}{-q-q_{\rm R}} \right \} \nonumber \\
&=& \frac{4 {\rm i} \gamma^2 \sqrt{s_{\rm r}} \, q}{(q^2-q^2_{\rm R})(q^2-q^{*2}_{\rm R})} \equiv 2 {\rm i} \Sigma(q) \label{eq:fRdisc}
\eea 
where 
\be 
q_{\rm R}^2 = q_0^2 - \gamma^2 - 2 {\rm i} q_0 \gamma \label{eq:qRsq}
\ee
and 
\be 
\Sigma(q) = \frac{2 \gamma^2 \sqrt{s_{\rm r}}\,  q}{(q^2 - q^2_{\rm R})(q^2 - q^{*2}_{\rm R})}.\label{eq:Sigmaq}
\ee 
Then according to (\ref{eq:fdr}) it should be the case that 
\be 
f_{\rm R}(s+{\rm i} \epsilon) = \frac{1}{\pi} \int_0^\infty {\rm d} q^{\prime 2} \frac{\Sigma(q')}{q^{\prime 2} - q^2 -{\rm i} \epsilon}. \label{eq:fRrec}
\ee 
The reader may verify by contour integration that the right hand side of (\ref{eq:fRrec}) does indeed reconstruct $f_{\rm R}$. 

The integration in (\ref{eq:fRrec}) is understood to be along the top side of the $s' \geq 4$ cut. It will be convenient in section 5.4.2 to consider the integration to be running just below the cut instead. 
In that case, $q'$ will be replaced by $-q'$ and we will have the representation 
\be 
f_{\rm R}(s_+) = \frac{1}{\pi} \int_{4, {\rm below}}^\infty {\rm d} s' \frac{\Sigma(s')}{s-s'+{\rm i} \epsilon} \label{eq:fRb}
\ee 
where 
\be 
\Sigma(s)= \frac{16 \gamma^2 \sqrt{s_{\rm r}}\sqrt{(s-4)}}{(s-I^2)(s-I^{*2})} \label{eq:Sigmas} 
\ee 
and 
\be 
I^2 = 4 + 4 q^2_{\rm R}. 
\ee 
The contour for (\ref{eq:fRb}) is shown in figure 12, which also exhibits the poles of the discontinuity function $\Sigma(s')$. 
\begin{figure}
\begin{center}
\includegraphics{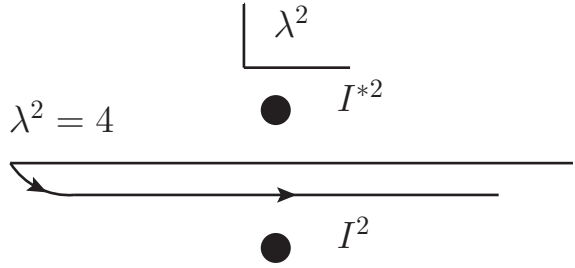} 
\end{center} 
\caption{The contour in the $\lambda^2$-plane for the integration in (\ref{eq:fRb}).}
\end{figure} 

\subsubsection{Left hand cut, the $N$ and $D$ functions}

Actually,
 amplitudes have other singularities in addition to those generated by unitarity - in particular, partial wave amplitudes have ``left hand'' singularities associated with exchange processes, which typically produce cuts along the real $s$ axis for $s \leq s_{\rm L}$, where 
$s_{L} < 4$. We can then write 
\be 
T(s) = \frac{{\rm e} ^{{\rm i} \delta} \sin \delta }{\rho} = \frac{N(s)}{D(s)}    \label{eq:NoverD}
\ee 
where $N(s)$ has only the LH cut and $D(s)$ has only the RH cut. In that case, 
\be
D_+ - D_- = N (T^{-1}_+ - T^{-1}_-) = - 2 {\rm i} \rho N  \label{eq: discD}
\ee
and we can therefore write (compare (\ref{eq:T-1dr})), 
\be 
D(s) = 1 - \frac{1}{\pi} \int_4^\infty \frac{\rho(s') N(s') {\rm d} s'}{s'-s}  \label{eq:Ddr} 
\ee 
assuming that $N(s)$ is such that the integral converges. 

Another representation for $D$ is sometimes useful. From (\ref{eq: discD}) we obtain 
\be 
D_+ - D_-=-2{\rm i} \, D_+ {\rm e}^{{\rm i} \delta} \sin \delta \label{eq:discDdelta}
\ee 
using (\ref{eq:NoverD}), so that 
\be 
D_+ = D_- {\rm e}^{- 2 {\rm i} \delta} \equiv D_- S^{-1}\label{eq:DDS-1}
\ee 
where $S$ is the $1 \times 1$ $S$ - matrix: $S= 1+2 {\rm i} T \rho$. 
Taking the logarithm of the first equation in (\ref{eq:DDS-1}), we see that the discontinuity of $\ln D$ is $- 2 {\rm i} \delta$, and so we can write 
\be 
D(s) = {\rm exp}\{-\frac{1}{\pi}\int_4^\infty \frac{\delta(s')}{s'-s} {\rm d} s'\}, \label{eq:Ddelta}
\ee 
always assuming convergence.

We shall not pursue the calculation of $2 \to 2$ amplitudes any further here. Instead our aim will be to see how unitarity and analyticity can provide useful formulae for the analysis of hadronic final state interactions, going beyond the simple $K$-matrix methods so far discussed. 

\subsection{Two hadron final state interactions} 

Let's return to the one-channel final state interaction (f.s.i.) discontinuity relation, written out again, 
\be 
F_+ - F_- = 2 {\rm i} T_- \rho F_+ = 2 {\rm i} T_+ \rho F_-,  \label{eq:Fdisc} 
\ee 
which we previously arranged to satisfy in the $K$ - matrix / $P$ - vector formalism. This time we're gong to include analyticity, as in our discussion of $L(s)$. 

First note that we can write (\ref{eq:Fdisc}) as 
\be 
F_+ = (1 + 2 {\rm i}T_+ \rho) F_- =  S F_-  \label{eq:F+SF-}
\ee 
 But we also know from (\ref{eq:DDS-1}) that $S=D_-/D_+$. It follows that 
\be 
F_+D_+ - F_- D_- =0, \label{eq:FD}
\ee 
so that the function $FD$ has no branch point at $s=4$. Hence the unitarity constraint is satisfied by 
\be 
F(s) = C(s)/D(s) \label{eq:C/D}
\ee 
where $C(s)$ is any function regular at $s=4$, for example a polynomial.

Now 
suppose that $F(s)$ has a ``background'' term $B(s)$ which we want to include - for instance, a Deck-type production process; $B(s)$ is assumed to have only a LH cut. We would like to take account of both $B(s)$ and  the unitarity constraint, in a way consistent with analyticity. In this case we can satisfy our unitarity and analyticity constraints by writing 
\be 
F(s) = B(s) + \frac{1}{\pi} \int_4^\infty \frac{T^{*}(s') \rho(s') F(s')}{s'-s} \, {\rm d} s', \label{eq:FBdr}
\ee
which is an {\em integral equation} for $F(s)$. Remarkably, there is an exact solution of 
this equation, due to  Omn\`{e}s \cite{omnes} and  Muskhelishvili \cite{mush}. 

Consider the discontinuity of the quantity $D(F-B)$ across the elastic $s \geq 4$ cut:
\bea
{\rm disc} (DF - DB) &=& D_+F_+ - D_- F_- - (D_+ - D_-) B \nonumber \\
&=& - (D_+ - D_-) B = 2 {\rm i} \rho N B. 
\eea
We can therefore write 
\be 
D_+ (F_+ -B) = \frac{1}{\pi} \int_4^\infty \frac{\rho(s') N(s') B(s')}{s'-s - {\rm i} \epsilon} {\rm d} s' 
\ee 
or 
\be 
F(s) = B + \frac{1} {\pi D} \int_4^\infty 
\frac{\rho' N' B' }{ s' - s }{\rm d} s'  . \label{eq:OM} 
\ee 
This is the famous O-M  solution to our f.s.i. problem. 

There are three points to note immediately: 

1) If $B$ is a constant (and the integral in (\ref{eq:OM})  converges), then $F$ becomes simply $B/D$.

2)  We can always add to  (\ref{eq:OM}) any solution to the ``homogeneous'' version of (\ref{eq:FBdr}) - that is, the equation with $B(s)=0$. We already know that such a solution has the form $F(s)=C(s)/D(s)$ with $C(s)$ 
regular at $s=4$.  So we may write the general solution to our problem as 
\be 
F(s) = B(s) + \frac{1}{\pi D(s)} \int_4^\infty \frac{\rho' N' B' }{ s' - s - {\rm I} \epsilon} {\rm d} s' + \frac{C(s)}{D(s)}. \label{eq:gensol}
\ee 

3) By making use of the identity 
\be 
\frac{1}{s' - s - {\rm i} \epsilon} = \frac{{\rm P.V.}}{s' - s} + {\rm i} \pi \delta (s' - s)   \label{eq:dirac} 
\ee 
where ``P.V.'' stands for ``principal part'', and taking $N$ to be approximately constant, we can write (\ref{eq:gensol}) as 
\be 
F \approx B {\rm e}^{{\rm i} \delta} \cos \delta + \frac{{\rm e}^{{\rm i} \delta} \sin \delta}{\rho} \left[ R + \frac{{\rm P.V.}}{\pi} \int_4^\infty \frac{\rho' B' {\rm d} s'}{s' -s} \right] \label{eq:gensolP}
\ee 
which is the more sophisticated version of our $K$-matrix formula  that analyticity has bought us. The task of fitting (\ref{eq:gensolP}) to data is simplified by the fact that the principal value integral is independent of the 2-body scattering parameters. Expression (\ref{eq: discD}) was used by Bowler {\em et al.} \cite{bowler1} to model diffractive $a_1$ production, using a Deck amplitude for $B(s)$. 

As with the $K$-matrix approach, the foregoing can be extended to the case of several coupled 2-body channels, so that $T, N, D$ and $\rho$ become matrices in channel space, while $F$ and $B$ are vectors. 
We shall not discuss 2-body final states further here, but turn our attention now to our main topic, the problem of f.s.i. among three hadrons.

\section{Final State Interactions Among Three Hadrons}

\subsection{Kinematics, and the isobar model} 

The amplitude we are now concerned with can be represented by figure 10, in which a state of definite
\begin{figure}
\begin{center} 
\includegraphics{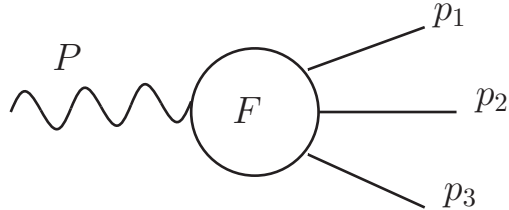} 
\end{center}
\caption{Three-hadron f.s.i. amplitude.}
\end{figure}
 $J^P$ decays to three hadrons. We first have to understand the kinematics of such decays. To simplify matters, we shall suppose that the three final state particles are spinless and of equal unit mass, while the initial state has $J^P = 0^+$ and invariant mass $m$ (i.e. this is the energy in the 3-body c.m. frame). 

Let the 4-momenta of the final state particles be $p_1, p_2$ and $p_3$, and let $P$ be that of the initial state, so that $P=p_1 + p_2 + p_3$, with $P^2 = m^2$. We introduce invariant variables $s, t, u$ by 
\be 
s=(p_2 + p_3)^2, t=(p_1+p_3)^2, u=(p_1 + p_2)^2 \label{eq:stu}
\ee 
which satisfy 
\be 
s+t+u=3+m^2. \label{eq:s+t+u}
\ee
Evaluating $t$ in the c.m.s. of particles 2 and 3 we find 
\be
t(s,x_s,m^2) = \frac{3+m^2-s}{2} - 2 p(s,m^2) q(s) x_s \label{eq:tsx} 
\ee 
where $x_s$ is the cosine of the angle between ${\bm p}_1$ and ${\bm p}_3$ in this system, and where 
\be 
p(s,m^2) = \{ [ s-(m-1)^2][s-(m+1)^2]\}^{1/2} /2 \sqrt{s} \equiv k(s, m^2)/2 \sqrt{s} \label{eq:psmsq}
\ee 
\be 
q(s) = (s-4)^{1/2}/2.  \label{eq:qdef}
\ee 
So $q(s)$ is the magnitude of the momentum of particle 2 or 3 in the 2-3 c.m.s., and $p(s, m^2)$ is the magnitude of the momentum of particle 1 in this system. The physical region for the decay process 
$m \to 1+2+3$ is then $4 \leq s \leq (m-1)^2$ and $|x_s| \leq 1$; the second condition can be written,  after some algebra, as 
\be 
stu -(m^2 -1)^2 \geq 0 \label{eq:pr} 
\ee 
or, using (\ref{eq:s+t+u})  as 
\be 
\G(s, t, m^2) \equiv st(3+m^2 - s -t) -(m^2-1)^2 \geq 0. 
\label{eq:kibble}
\ee 
$\G$ is the Kibble \cite{kibble} cubic, drawn in figure 14 for the case  $m >3$.  The physical region for the decay (the Dalitz region)  is inside the closed loop labelled ${\mathcal{D}}$. The regions labelled I, II and II are physical regions for the `crossed' reactions $m+{\bar{1}} \to 2+3 \, ({\rm I}),\,  m+{\bar{2}} \to 1+3 \, ({\rm II}), \,  m+{\bar{3}} \to 1 + 2 \, ({\rm III})$. 
\begin{figure}
\begin{center} 
\includegraphics[scale=.7]{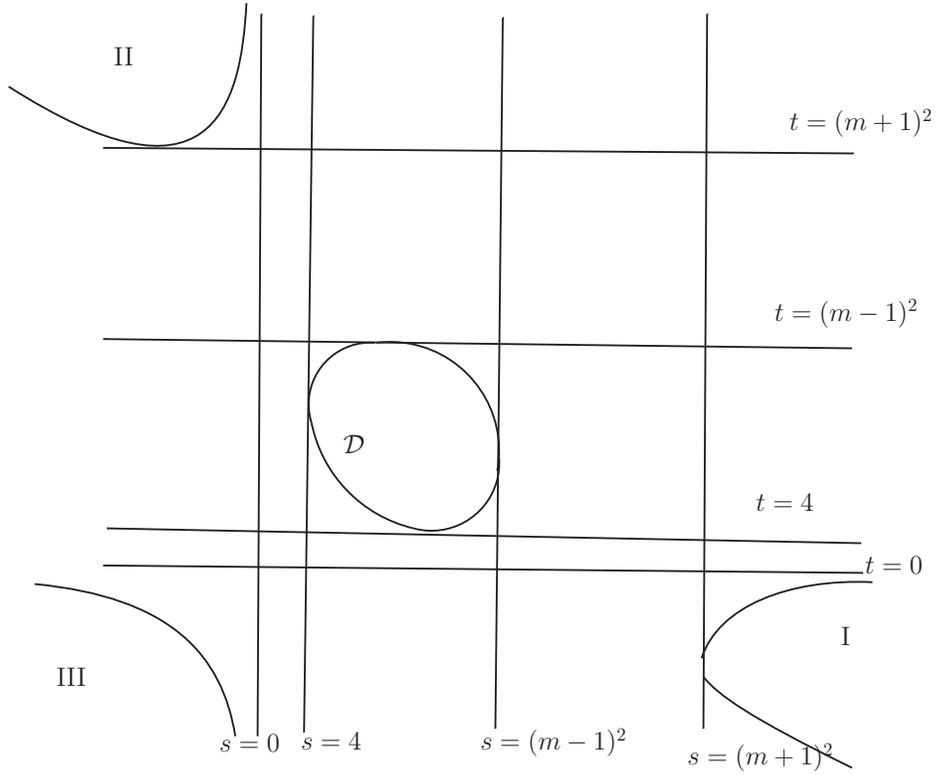}
\caption{The Kibble cubic, with the region for the decay $m \to 1+2+3$ labelled ${\mathcal{D}}$ (the Dalitz plot region).}
\end{center}
\end{figure} 

A figure showing how experimental events populate the Dalitz region is called a Dalitz plot - first invented by Dalitz \cite{dplot} in connection with his famous analysis of $K \to \pi \pi \pi$ (which led to the discovery of parity violation). If the matrix element for the decay is a constant, then the events will be uniformly distributed on the plot. This follows from the fact that the dependence on $s$ and $t$ in the three-particle phase space is proportional to 
\be 
\frac{{\rm d}s {\rm d} t} {m^2}.   \label{eq:3ps} 
\ee 
If, on the other hand, a 2-particle resonance can be formed in any of the three pairs of particles, then there will be strong concentrations of events along bands centred on the square of the resonance mass(es). 

It is an empirical fact, of course, that very many three-hadron systems are such that their two-hadron subsystems do indeed form resonances (e.g. $\pi^+ \pi^- \pi^+, K^+ \pi^- \pi^+, $ etc.) This fact is the basis for the {\em isobar model} \cite{isobar}, which expresses the $m \to 1+2+3$ decay amplitude as a coherent superposition of ``resonance + spectator particle'' states, as in figure 15. In practical
\begin{figure}
\begin{center}
\includegraphics[scale=.7]{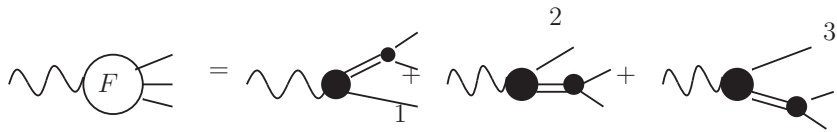}
\end{center}
\caption{Isobar model expansion of $F$ as a coherent superposition of resonance + particle amplitudes.}
\end{figure}
 applications, one has to deal seriously with all the complications of spin, angular momentum, recoupling coefficients etc. Eventually we shall come back to these complications, briefly, but for the most part we shall for pedagogical purposes concentrate on the simplest model, in which all three identical particles are spinless, each pair forms a resonance in the $l=0$ state, and the overall $J^P = 0^+$. For this toy isobar model, then, referring to figure 15  we write 
\be 
F(s,t,u,m^2) = C(m^2)M(s) + C(m^2) M(t) + C(m^2) M(u)  \label{eq:Fisobar} 
\ee 
where $C(m^2)$ is a  common ``production vertex'' represented by the solid blob, and  depends on 
$m^2$ but not on $s$, while $M(s)$ is the elastic $2 \to 2$ amplitude in the 2+3 channel having only the $s \geq 4$ cut (and similarly for $M(t) $ and $M(u)$). The factorisation of each term in (\eqref{eq:Fisobar}) into a product of a function of $m^2$ times a function of $s$ is fundamental to the isobar model - and is also, as we shall soon see, inconsistent with unitarity.

\subsection{The isobar model violates subenergy unitarity} 

The constraint which determines the correct phase in a subenergy channel is the subenergy discontinuity (unitarity) relation, shown diagrammatically in figure 16 for the $s$ channel. This is very similar to figure
\begin{figure}
\begin{center}
\includegraphics{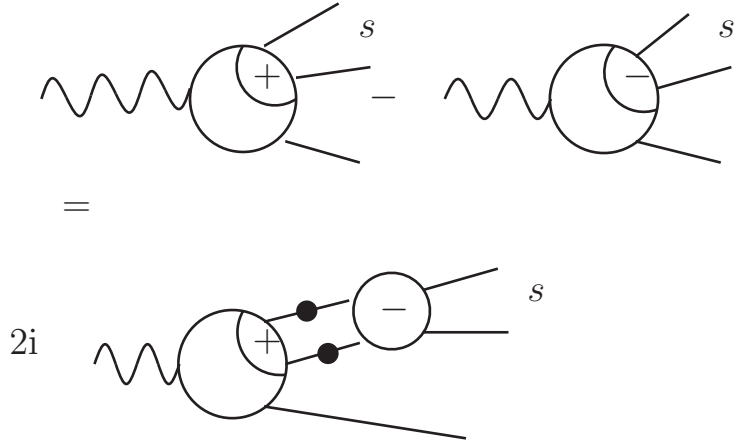}
\end{center}
\caption{Subenergy discontinuity relation.}
\end{figure}
 9, but with one crucial difference: the amplitude $F(s,t,u,m^2)$ now contains (unlike $F(s)$ of figure 9) an angle dependence via (\ref{eq:tsx}), which must be integrated over in the two-particle phase space integral:
\bea 
\lefteqn{s \geq 4:F(s_+, t, u, m^2) - F(s_-, t, u, m^2) = } \nonumber \\
 & & 2 {\rm i} \rho(s) \frac{1}{2} \int_{-1}^1 {\rm d} x_s F(s_+, t(s_+, x_s), u(s_+, x_s), m^2) M(s_-). \label{eq:subendisc} 
\eea
Equation (\ref{eq:subendisc}) is the required generalisation of (\ref{eq:discF}). Recall  that our  normalization for $\rho(s)$ is 
\be
\rho(s)=\left( \frac{s-4}{s}\right)^{1/2}.
\ee

It is quite simple to see that (\ref{eq:Fisobar}) cannot satisfy (\ref{eq:subendisc}). By construction, $M(t)$ and $M(u)$ have only the cuts $t \geq 4, u \geq 4$, and no discontinuity in $s$. So the LHS of (\ref{eq:subendisc}) is just 
\be 
C(m^2) (M(s_+) - M(s_-)) = C(m^2) 2 {\rm i} \rho (s) M_+(s) M_-(s).
\ee 
 This is the same as the RHS of 
(\ref{eq:subendisc}) only if the $M(t)$ and $M(u)$ terms in $F$ are absent. 

The expansion (\ref{eq:Fisobar}) therefore has to be modified in order for $F$ to satisfy (\ref{eq:subendisc}). An economical way to do this is to replace $M(s)$ in (\ref{eq:Fisobar}) 
by $M(s)\phi(s,m^2)$ (and similarly for $M(t)$ and $M(u)$), where $\phi(s,m^2)$ is going to be determined from (\ref{eq:subendisc}) and analyticity. We have anticipated, and will soon confirm, that the correction function $\phi$ will depend on both $s$ and $m^2$, and will thus spoil the factorisation 
in (\ref{eq:Fisobar}) which was alluded to earlier. 

So we now take 
\be 
F(s,t,u,m^2) = C(m^2) [ M(s) \phi(s, m^2) + M(t) \phi(t, m^2) + M(u) \phi(u,m^2)]. \label{eq:Fcorr}
\ee
Inserting (\ref{eq:Fcorr}) into (\ref{eq:subendisc}) the LHS is (in shortened notation) 
\bea
C (M_+\phi_+  - M_-\phi_-) &=& C ((M_+ - M_-) \phi_+ + M_- (\phi_+ - \phi_-)) \nonumber \\
&=& C (2 {\rm i} \rho M_+ M_- \phi_+ + M_- (\phi_+ - \phi_-))  \label{eq:corrdisc}
\eea 
while the RHS is 
\be 
2 {\rm i}\,  \rho\, C \, M_+\, M_- \,  \phi_+ + 2 {\rm i}\, \rho\, 2 C \,M_- \, \frac{1}{2}\, \int_{-1}^1 M(t) \phi(t, m^2){\rm d} x_s \label{eq"corrdisc2} 
\ee 
since the two contributions from the $t$ and $u$ terms are equal. It follows that 
\be
s \geq4: \,\phi(s_+, m^2) - \phi(s_-, m^2) = 2 {\rm i} \rho(s_+) \int_{-1}^1 M(t) \phi(t, m^2){\rm d} x_s. \label{eq:discphi}
\ee
This important equation tells us that the function $\phi(s,m^2)$ has a discontinuity across the $s \geq 4$ cut, determined to be (\ref{eq:discphi}) by unitarity in the $s$-channel. It is evident from (\ref{eq:discphi}) that $\phi$ must depend on $m^2$ as well as on $s$, via the $m^2$-dependence of $t$ 
(c.f. (\ref{eq:tsx})). Equation (\ref{eq:discphi}) also implies that $\phi$ will develop an imaginary part for $s \geq 4$, which means that the phases carried by the terms $M \phi$ in (\ref{eq:Fcorr}) are no longer those carried by the two-body amplitudes $M$ in (\ref{eq:Fisobar}). The importance of examining the constraint of subenergy unitarity was especially emphasized by Aaron and Amado \cite{AA}. 

\subsection{Implementing  subenergy unitarity and analyticity}

\subsubsection{Failure of a ``$K$-matrix'' approach}

We might be tempted to implement the subenergy unitarity constraint by a $K$-matrix type of procedure, exploiting the fact (as before) that $\rho_+ - \rho_- = 2 \rho$. We would then write 
\be 
\phi(s,m^2) \stackrel{\textstyle{?}}{=} ({\rm regular \ function}) + {\rm i} \rho \int_{-1}^1 M(t) \phi(t, m^2) {\rm d} x_s.  \label{eq:phiK} 
\ee
Unfortunately, though simple, this prescription is incorrect, in the sense that it leads to singularities in $\phi$ which perturbation theory teaches us should not be impacting the physical region. To see this, note first that the uncorrected isobar model corresponds to taking $\phi=1$. So we will make (\ref{eq:phiK}) more precise by writing 
\be 
\phi(s, m^2) \stackrel{\textstyle{?}}{=}  1 + {\rm i} \rho \int_{-1}^1 M(t) \phi(t, m^2) {\rm d} x_s  \label{eq:phiK1}
\ee 
and imagine solving this integral equation iteratively. The first correction will be 
\be 
J(s, m^2) = {\rm i} \rho \int_{-1}^1 M(t(s,x_s)) {\rm d} x_s.   \label{eq:itKphi}
\ee

Suppose that $M$ has one resonance, which for the moment we may parametrise as\footnote{This form ignores the normal threshold branch point at $t=4$  present in $f(t)$ given by (\ref{eq:BW}) or in $f_{\rm R}(t)$ of (\ref{eq:BWq}). We will treat this properly in section 5.4.}  
\be
M(t(s,x_s)) = \frac{4 \gamma \sqrt{s_{\rm r}}}{I^2 - t(s,x_s) }, \ \ I^2=m_{\rm R}^2 - {\rm i} \Gamma.    \label{eq:MBW}
\ee
The denominator is a linear function of $x_s$, and the integral is easily done yielding the  
result $J(s, m^2)= 2{\rm i} \rho(s) B(s, m^2)$ where 
\be 
B(s, m^2) = \frac{4 \gamma \sqrt{s_{\rm r}}}{4 p(s , m^2) q(s)} \ln \frac{m^2_{\rm R} - t(s, x_s=+1) - {\rm i} \G}{m^2_{\rm R} - t(s, x_s=-1)- {\rm i} \G}.  \label{eq:J}
\ee
The logarithm has singularities in $s$ when 
\be 
t(s, x_s= \pm 1) = m_{\rm R}^2 - {\rm i} \G. \label{eq:singJ}
\ee 
The two curves $t(s, x_s=+1)$ and $t(s, x_s=-1)$ together form the boundary of the $s-t$ Dalitz plot as $s$ varies. The singularity at $t(s, x_s=+1) = m^2_{\rm R} -{\rm i} \Gamma$ occurs at that value of $s$, say $s_b$,  at which the boundary arc $t(s, x_s=+1)$ hits the resonance band centred at $t=m^2_{\rm R}$; similarly for the other singularity, at $s_a$. 

Though ``only'' logarithms, these singularities cause quite noticeable phase and modulus variation - but are they to be believed? The answer is no: these singularities are largely spurious \cite{spurious}. They are in fact the positions of singularities (in $s$) of the triangle graph shown in figure 17. Such diagrams have been thoroughly discussed (and we shall soon meet them). One
\begin{figure}
\begin{center} 
\includegraphics{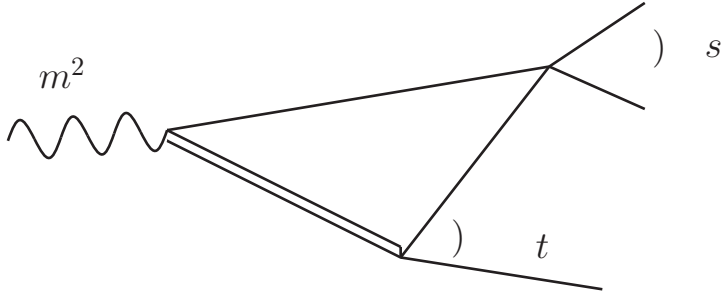}
\end{center} 
\caption{First rescattering correction, or triangle graph.}
\end{figure} 
singularity, $s_b$, is near threshold and can be near the physical sheet of $\phi$, but its proximity to threshold limits its effect. The other, at $s_a$, is far from the physical region. 

\subsubsection{Combining unitarity and analyticity}

Why are singularities openly present in $B(s, m^2)$ somehow masked in the triangle graph? 
Just as in the case of the $s=0$ singularity of $\rho(s)$, we need to include {\em analyticity}, in addition to unitarity, in order to get a physically correct amplitude. In other words, we have to insert the discontinuity relation (\ref{eq:discphi}) into a dispersion relation. Singularities present in the integrand can get moved away from the physical region after integration.  

This leads immediately to the equation 
\be 
\phi(s, m^2) = 1 + \frac{2}{\pi} \int_4^\infty \frac{\rho(s')}{s'-s-{\rm i} \epsilon} 
\{  \frac{1}{2} \int_{-1}^1 {\rm d} x_s M(t(s', x_s)) \phi(t(s', x_s), m^2) \} {\rm d} s' 
\label{eq:phidr} 
\ee 
where we are assuming that the integral over $s'$ will converge, and are also taking the inhomogeneous term to be unity, corresponding to the unmodified isobar model, as in (\ref{eq:phiK1}). It will sometimes be convenient to define $\Phi(s, m^2) = M(s) \phi(s)$ and rewrite (\ref{eq:phidr}) as 
\be 
\Phi(s,m^2) = M(s) + 2 M(s) \frac{1}{\pi} \int_4^\infty \frac{\rho(s')}{s'-s-{\rm i} \epsilon} \{ \frac{1}{2} \int_{-1}^1 \Phi(t(s', x_s)) {\rm d} x_s \} {\rm d} s'.   \label{eq:Phidr}
\ee
Equations (\ref{eq:phidr}) and (\ref{eq:Phidr}) are integral equations embodying the basic constraints of two-body unitarity and analyticity. They are therefore a kind of minimal theory of corrections to the isobar model, which would correspond to just the first term in (\ref{eq:phidr}) and (\ref{eq:Phidr}). Equations of this type were first proposed by Khuri and Treiman \cite{KT}, from a rather different standpoint. 

\subsection{The first rescattering correction: the triangle graph}

\subsubsection{The rescattering amplitude $T(s, m^2)$}

It is possible to proceed directly on the basis of (\ref{eq:Phidr}), solving it iteratively, for example. The first iteration is again just the usual two-body amplitude used in the isobar model, and the first correction to this adds to it   the amplitude 
\be 
\frac{2 M(s)}{\pi} \int_4^\infty \frac{\rho(s')}{s' - s - {\rm i} \epsilon} \{ \frac{1}{2} \int_{-1}^1 M(t(s', x_s)) {\rm d} x_s\}. \label{eq:firstit} 
\ee 
A particularly interesting case is that in which the amplitude $M(t)$ is resonant. It is important to 
make sure that we are getting the sheet structure of $M$ correct, so we will set $M(t) = f_{\rm R}(t)$ where $f_{\rm R}(t)$ has the representation (\ref{eq:fRb}): 
\be 
M(t)= \frac{1}{\pi} \int_{4, {\rm below}}^\infty {\rm d} \lambda^2 \frac{\Sigma(\lambda^2)}{t-\lambda^2 + {\rm i} \epsilon} \label{eq:MfR}
\ee
and 
\be 
\Sigma(\lambda^2) = \frac{16 \gamma^2 \sqrt{s_{\rm r}} \sqrt{s-4}}{(\lambda^2-I^2)(\lambda^2-I^{*2})} 
\label{eq:Sigmalam}
\ee
with 
\be 
I^2 = 4+4(q_0^2-\gamma^2-2{\rm i}q_0\gamma). \label{eq:Isq}
\ee
The first iteration of (\ref{eq:Phidr}) then gives 
\be 
\Phi^{(1)}(s, m^2) = M(s) [1 + T(s, m^2)] \label{eq:Phi1}
\ee 
where 
\be 
T(s, m^2) = \frac{1}{\pi} \int_4^\infty {\rm d} s' \frac{\rho(s')}{s'-s - {\rm i} \epsilon}
\frac{1}{2} \int_{-1}^{+1}{\rm d} x_s  \frac{1}{\pi} \int_{4, {\rm below}}^\infty {\rm d} \lambda^2 \frac{\Sigma(\lambda^2)}{t - \lambda^2 + {\rm i} \epsilon} \label{eq:Tsmsq}
\ee
is the {\em first rescattering correction}. We have dropped the factor of 2 since we want the contribution from just one rescattering channel, and we have also suppressed the (important) arguments 
$s', x_s$ of $t$.  We may rewrite (\ref{eq:Tsmsq}) as 
\be 
T(s, m^2) = - \frac{1}{\pi} \int_{4, {\rm below}}^\infty \Sigma(\lambda^2) f(s, \lambda^2, m^2)
\label{eq:Tsmsq1}
\ee 
where 
\be 
f(s, \lambda^2, m^2)=  \frac{1}{\pi} \int_4^\infty {\rm d} s' \frac{\rho(s')}{s' - s -{\rm i} \epsilon} \, \frac{1}{2} \int_{-1}^{1} {\rm d} x_s \frac{1}{\lambda^2 - t - {\rm i} \epsilon}
 \label{eq:fslm}
 \ee
is the triangle graph of figure 18 with two internal particles of unit mass and one of squared  mass 
$\lambda^2$. 
\begin{figure} 
\begin{center} 
\includegraphics{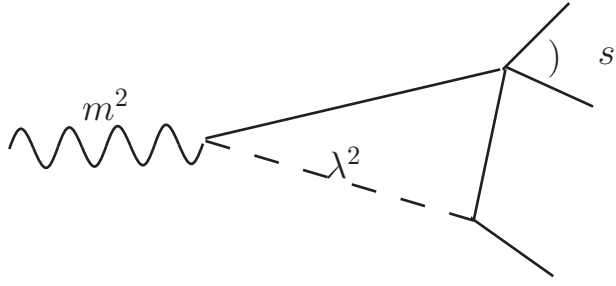} 
\end{center} 
\caption{The triangle graph amplitude of (\ref{eq:fslm}).}
\end{figure}

We can understand this graphical interpretation   by considering how we would calculate this diagram by writing a dispersion relation in the variable $s$. Looking along the $s$-channel, we see a normal threshold at $s=4$, with a discontinuity given by cutting the graph across the two internal lines  of unit mass, which are put on mass-shell. This discontinuity is proportional to the product of the two-body phase space factor $\rho(s)$ and the $s$-wave projection (in the $s$- channel c.m.s.) of the $t$-channel exchange diagram, as  shown in figure 19. These are exactly the ingredients of (\ref{eq:fslm}).
\begin{figure}
\begin{center} 
\includegraphics{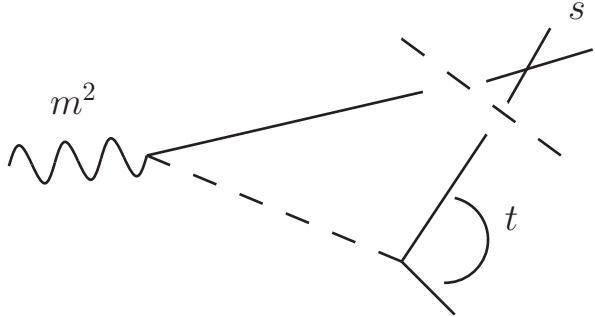} 
\end{center} 
\caption{Reconstructing figure 18 by dispersing in $s$.}
\end{figure} 

The amplitude $T(s, m^2)$ of (\ref{eq:Tsmsq1}) is therefore an integral over the variable internal squared mass $\lambda^2$ of the  triangle graph, weighted by the spectral function $\Sigma(\lambda^2)$. We may represent $T(s, m^2)$ by figure 20. 
\begin{figure}
\begin{center} 
\includegraphics{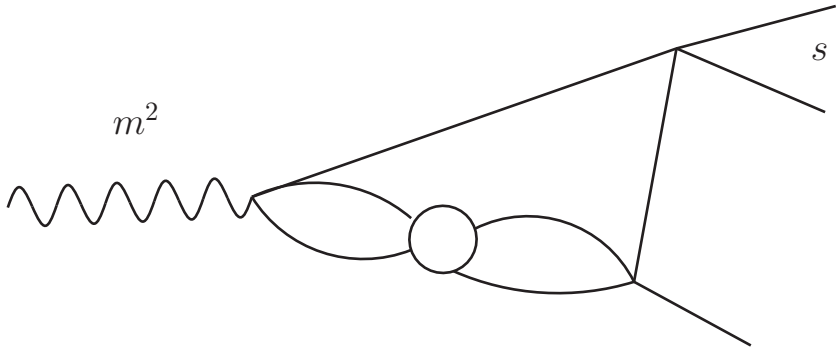} 
\end{center} 
\caption{The first rescattering correction $T(s, m^2)$.}
\end{figure} 

\subsubsection{Singularities of $T(s, m^2)$ near the physical region}

 An important question is whether $T(s, m^2)$ as given by (\ref{eq:Tsmsq1}) has any singularities in $s$ or $m^2$ which are near the physical region in those variables, since they would be  likely to cause significant  variation in the magnitude and phase of $T(s, m^2)$. The singularities of amplitudes such as $T(s, m^2)$ were studied in \cite{AK1}, whose analysis we now briefly describe. .  

 It is clear first of all that $T$ has the normal threshold branch point at $s=4$, since this is present in $f(s, \lambda^2, m^2)$. As usual, this results in a two-sheeted structure for $T$. The physical amplitude $T(s, m^2)$ is obtained from (\ref{eq:Tsmsq1}) by integrating, with $s$ approaching the real axis from above, the physical sheet amplitude of $f(s, \lambda^2, m^2)$ along a contour taken just below the real $\lambda^2$ axis, as shown in figure 12. Also in this figure we have indicated the positions of the poles in $\Sigma(\lambda^2)$ at $\lambda^2 = I^2$ and $ \lambda^2 = I^{*2}$. 
 
 There is another way in which a singularity of $T(s, m^2)$ can be generated. It can be shown \cite{BK} 
 that $f(s, \lambda^2, m^2)$ has two singularities at $\lambda^2 = \lambda^2_{\pm}(s, m^2)$ which move around in the $\lambda^2$-plane as $s$ (or $m^2$) move. It may happen that, as $s$ moves in the complex $s$-plane, one of these singularities - say $\lambda^2_+(s, m^2)$ - approaches the $\lambda^2$ contour from above, so that the contour has to be deformed away from the advancing singularity in order to have a smooth continuation, as shown in figure 21. If it should happen that the advancing $\lambda^2_+(s, m^2)$ actually pins the contour against the pole of $\Sigma(\lambda^2)$ at $\lambda^2=I^2$, so that the contour cannot be deformed away, then for that value of $s$ and $m^2$ there will be a singularity of $T(s, m^2)$. This is called a ``pinch'' singularity, for obvious reasons. 
 \begin{figure}
 \begin{center} 
 \includegraphics{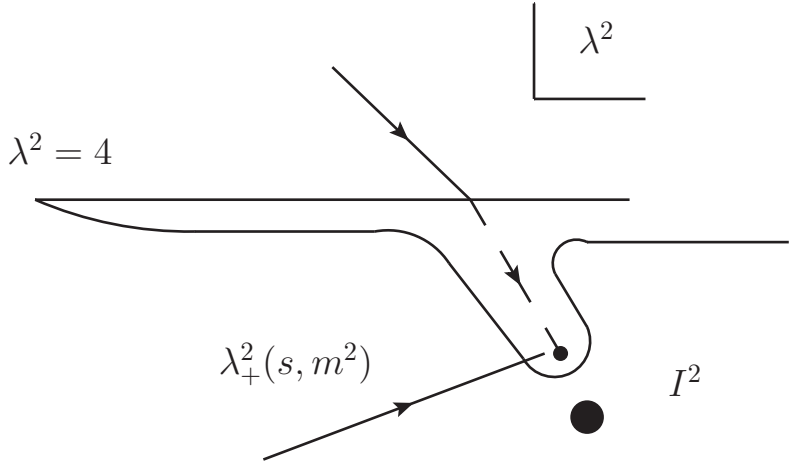} 
 \end{center} 
 \caption{Pinch singularity in $T(s, m^2)$.}
 \end{figure} 
 
 Careful analysis \cite{BK} \cite{AK1} shows that it is possible for such a pinch singularity of $T(s, m^2)$   to occur at a point $s=s_b(m^2)$, near the physical region in $s$. The singularity, which is logarithmic,  is present on the second $s$-sheet of $T(s, m^2)$, reached as usual by crossing the real axis from just above the $s \geq 4$ cut. The imaginary part of $s_b$ is related to that of $I^2$, and for a narrow resonance $s_b$ will be close to the real axis and therefore near the physical region for $T(s, m^2)$. 
 
 This situation only arises for a particular range of values of $s$ and $I^2$, for fixed $m^2$. This range is easily visualized on a Dalitz plot for the variables $\lambda^2$ and $s$. Referring to figure 22, the pole at $\lambda^2=I^2$ is represented as a resonance band at fixed $\lambda^2$. This band intersects the boundary of the plot at two points $s_b$ and $s_a$ (assuming the imaginary part of $I^2$ is small). Both of these points are potentially singularities of $T(s, m^2)$ - indeed they are just the same points as those we encountered  using the (incorrect) ``$K$-matrix'' unitarisation scheme. Here,  analysis shows \cite{AK1} that $s_a$ is never near the physical region and $s_b$ is only near it if $I^2$ lies in the range 
 \be 
 \frac{1}{2}(m^2-1) \leq I^2 \leq (m-1)^2, \label{eq:Isqrange}
 \ee 
 neglecting the complex part of $I^2$. The corresponding $s_b$ lies in the range 
 \be 
 4 \leq s_b \leq (m+1).\label{eq:sbrange}
 \ee 
 The ranges (\ref{eq:Isqrange}) and (\ref{eq:sbrange}) are of course given for our current (unit) mass values. In general, the resonance band must intersect the Dalitz plot boundary on the upper left hand arc, and the singularity $s_b$ is read off on the $s$-axis from this intersection. 
 \begin{figure}
 \begin{center} 
 \includegraphics{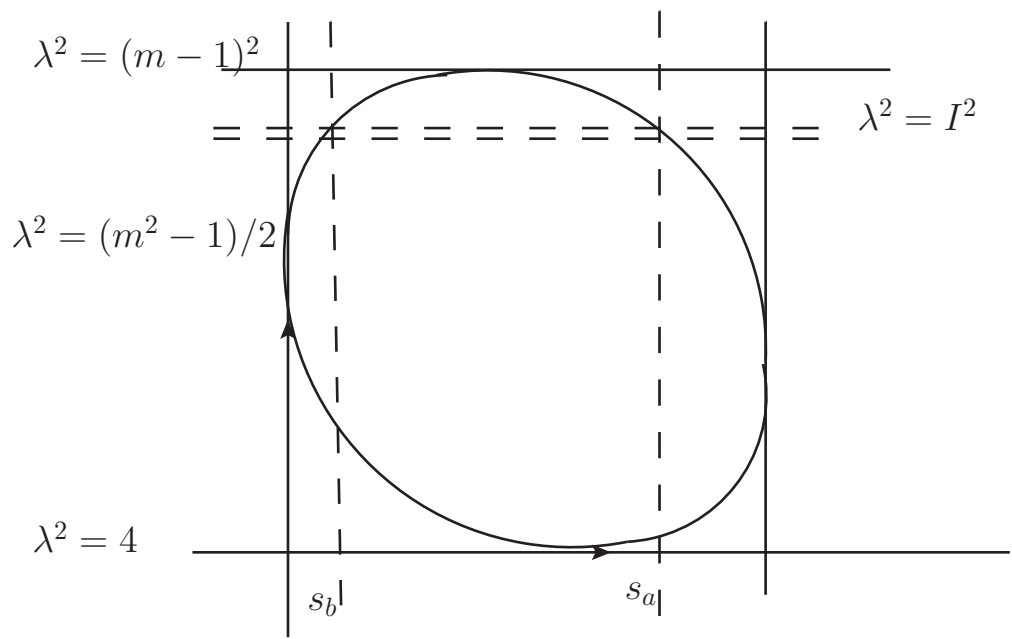} 
 \end{center} 
 \caption{The situation in which the triangle singularity $s_b$ is near the physical region.}
 \end{figure}

 The upper limit of the range (\ref{eq:Isqrange}) corresponds to the $m^2$ value $m^2=(I+1)^2$, which is just the ``normal threshold'' for making the ``quasi two-particle'' state consisting of one particle of complex mass $I$ and another particle of unit mass. We shall discuss such particle + resonance states in section 8. It is clear that no nearby singularity associated with the resonance can occur unless $m^2$ is at least greater than the threshold value $(I+1)^2$. For this $m^2$, $s_b(m^2)$ is close to the point $m+1$, with a small negative imaginary part related to that of $I^2$. As $m^2$ increases, the Dalitz plot grows, and the intersection point $s_b(m^2)$ moves towards the point $s=4$, reaching it when $m^2=2I^2 +1$. Thereafter $s_b(m^2)$ moves around $s=4$ into the upper half plane, still on the second $s$-sheet, but progressively further from the physical region. This motion of $s_b(m^2)$ is shown in figure 23. In general, $s_b$ will lie close to threshold. 
  \begin{figure}
  \begin{center} 
  \includegraphics{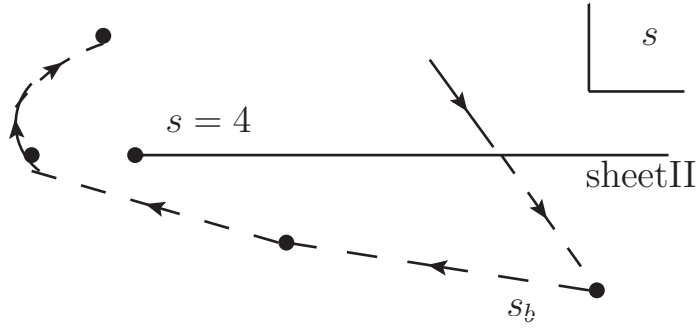} 
  \end{center} 
  \caption{The motion of $s_b(m^2)$ as $m^2$ increases from $(I+1)$ to greater than $2I^2+1$ .}
  \end{figure}

 The amplitude $T(s, m^2)$ could be evaluated directly from the representation (\ref{eq:Tsmsq1}). However, it would be somewhat more intuitive if we could somehow extract from the $\lambda^2$ integral  the contribution associated with the pole at $\lambda^2=I^2$. We would then, up to a constant factor,  be dealing with  $f(s, I^2, m^2)$, which is the triangle graph with an internal particle of squared mass $I^2$, shown in figure 17.  This can easily be done. Recall that the singularity at $s_b$ arises from a pinch of the $\lambda^2$ contour in figure 21 between the singularity at $\lambda^2=\lambda^2_+(s,m^2)$ of $f(s, \lambda^2, m^2)$ and the pole at $\lambda^2=I^2$. If the $\lambda^2$ contour passed {\em below} the pole, $\lambda^2_+(s, m^2)$ and the pole would be on the same side of the contour, and no pinch would occur. Let us denote the amplitude defined along such a contour by $\hat{T}(s, m^2)$. Then $\hat{T}$ is free of the singularity at $s_b$. Referring to figure 24, we see that this second contour
  \begin{figure}
   \begin{center} 
   \includegraphics[scale=.7]{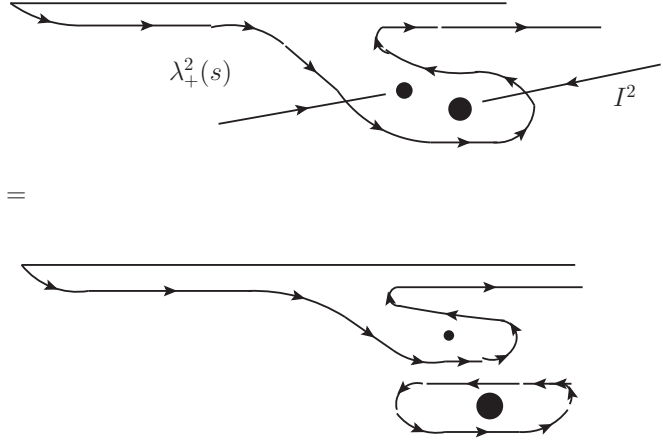} 
   \end{center} 
   \caption{The $\lambda^2$ contours for (\ref{eq:TThat}).}
   \end{figure} 
   (for $\hat{T}$) is equivalent to the first contour (for $T$) together with a circuit around the point $\lambda^2=I^2$. Hence 
 \be 
 T(s, m^2) = {\hat{T}}(s, m^2) + \frac{1}{\pi} 2 \pi {\rm i} R \, f(s, I^2, m^2) \label{eq:TThat}
 \ee 
 where $R$ is the residue of $\Sigma(\lambda^2)$ at the pole $\lambda^2=I^2$. The function $f(s, I^2, m^2)$ does contain the nearby singularity $s_b$, while  the function $\hat{T}(s, m^2)$ does not. We can therefore calculate the effect of the singularity $s_b$ by evaluating the quantity $2 {\rm i} R \, f(s, \lambda^2, m^2)$. Since $R=2 {\rm i} \gamma \sqrt{q^2_{\rm R}}/q_0$, we obtain finally for the singular part of the rescattering amplitude 
 \be 
 T_{\rm sing} (s, m^2) = - 4 \gamma \sqrt{s_{\rm r}} \sqrt{q^2_{\rm R}}/q_0 \, f(s, I^2, m^2). \label{eq:Tsing}
 \ee
 
 Noting now that $\sqrt{q^2_{\rm R}}=(q_0^2 - \gamma^2 - 2 {\rm i} q_0 \gamma)^{1/2} \approx -q_0$, we see that we have arrived at the reassuring result that $T_{\rm sing}$ is, to a good approximation,  just the triangle graph of figure 17, using the ``naive'' amplitude $4 \gamma \sqrt{s_{\rm r}}/(I^2-t)$ for the $t$-channel resonance (i.e. ignoring the branch point at $t=4$):
 \bea 
 T_{\rm sing}(s, m^2) &\approx& \frac{1}{\pi} \int_4^\infty {\rm d} s' \frac{\rho(s')}{s'-s-{\rm i} \epsilon} \, \frac{1}{2} \int_{-1}^1 \frac{4 \gamma \sqrt{s_{\rm r}}}{I^2-t} {\rm d} x \nonumber \\
 &=& \frac{1}{\pi} \int_4^\infty {\rm d} s' \frac{\rho(s')}{s'-s - {\rm i} \epsilon} B(s', m^2) \label{Tsingap}
 \eea
 where $B(s, m^2)$ is given in (\ref{eq:J}).
 
 \subsubsection{Physical picture of the nearby rescattering singularity}
 
 In the limit where the imaginary part of $I^2$ goes to zero, so the resonance has zero width, the singularity $s_b(m^2)$ approaches the real axis. This is an infinity in  $f(s, I^2, m^2)$ in the physical region (though not of $T$, in which $f$ is multiplied by the width parameter $\gamma$). How can such a physical region singularity of a Feynman graph occur? The general answer was provided in an elegant paper by Coleman and Norton \cite{CN}. They showed that a Feynman amplitude has singularities in the physical region if and only if the corresponding Feynman diagram can be interpreted as a picture of a four-momentum conserving process occurring in space-time, with all internal particles on-shell, and moving forward in time. The particular case of this result for the triangle diagram was given by Bronzan 
 \cite{JBB}. 
 
 Following \cite{JBB} for our simple case of three identical spinless particles of unit mass, and a resonance of squared mass $I^2$, consider the rescattering graph of figure 17, in which the resonance is in the (13) or $t$-channel, and the final rescattering is in the (23) or $s$-channel. For this to be a real physical process, we certainly need $m^2 \geq (I+1)^2$, or $I^2 \leq (m-1)^2$ as in (\ref{eq:Isqrange}). The lower inequality in (\ref{eq:Isqrange}) arises from the ``catch-up'' condition: namely, in the rest frame of $I^2$, the decay particle 3 must be moving in the same direction as the 
 ``fleeing'' particle 2, and the speed of particle 3 must be greater than or equal to the speed of particle 2. The kinematics is similar to that in section 5.1, except that now we work in the (13) c.m.s. rather that the (23) c.m.s., and we set $t=I^2$. So we write (c.f. (\ref{eq:tsx})-(\ref{eq:qdef}))
 \be 
 s = (3+m^2 -t)/2 - 2 p(t, m^2) q(t) x_t   \label{eq:tcmskin}
 \ee 
 where 
 \be p(t, m^2) = [t^2-2t(m^2+1) + (m^2-1)^2] ^{1/2}/2 \sqrt{t} \label{eq:ptm}
 \ee 
 and 
 \be 
 q(t) = (t-4)^{1/2}/2,
 \ee
 and $x_t$ is the cosine of the angle between 1 and 3 in this $t$-channel c.m.s. The speed of particle 3 is then $[(I^2-4)/I^2]^{1/2}$. The energy of particle 2 is $(m^2 -I^2-1)/2\sqrt{I^2}$, and the magnitude of its momentum is $p(I^2, m^2)$. We therefore require, for a physical rescattering, $x_t=+1$ and 
 \be 
 [(I^2-4)/I^2]^{1/2} \geq \frac{[I^4-2I^2(m^2+1) +(m^2-1)^2]^{1/2}}{m^2-I^2-1} \label{eq:v3gtv2}
 \ee
 which reduces to 
 \be 
 I^2 \geq (m^2-1)/2  \label{eq:Isqll}
 \ee 
 as in the lower inequality of (\ref{eq:Isqrange}). The arc $x_t=+1$ is shown in figure 25, from which, together with (\ref{eq:Isqll}) we see that the catch-up conditions are precisely that the $I^2$ band intersects the Dalitz boundary on the upper left-hand arc AB. This condition is completely general, for arbitrary mass values in the triangle graph. 
  \begin{figure}
    \begin{center} 
    \includegraphics{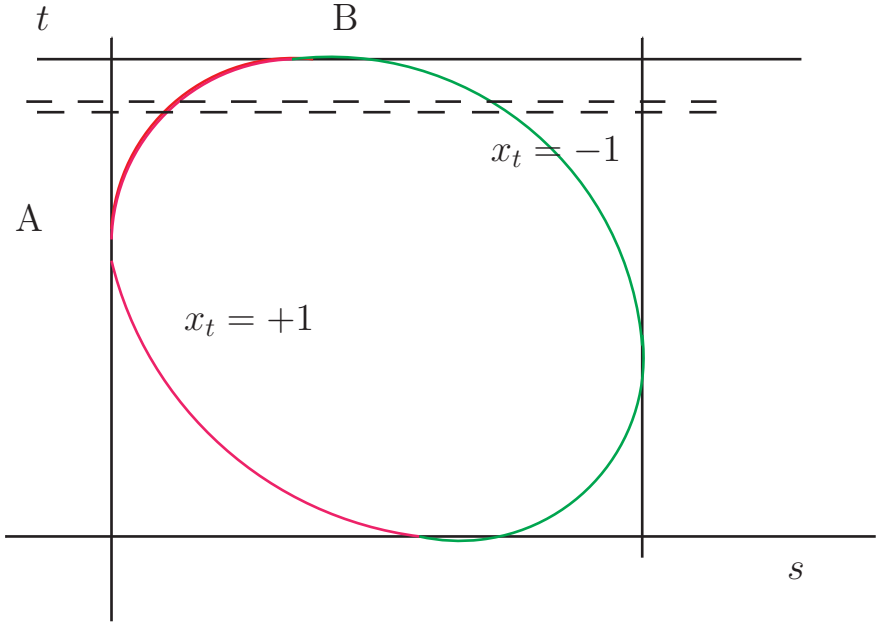} 
    \end{center} 
    \caption{The region AB of the Dalitz plot boundary where the rescattering can occur as a physical process. }
    \end{figure} 
 
 \subsubsection{Some examples}
 
 Under what circumstances might a nearby singularity $s_b$ be potentially observable? We'll return to this question in the next section, but first we discuss some possible examples. 
 
 In the case of three identical final state particles, there will be a resonance at $s=I^2$ in $M(s)$, and this amplitude multiplies $f(s, I^2, m^2)$ in (\ref{eq: discD}). This situation is shown in figure 26. It is clear that $s_b$ will lie far from the region where $M(s)$ is large, and the net effect will be only a small modification of the tail of the resonance in $M(s)$. 
  \begin{figure}
    \begin{center} 
    \includegraphics[scale=.8]{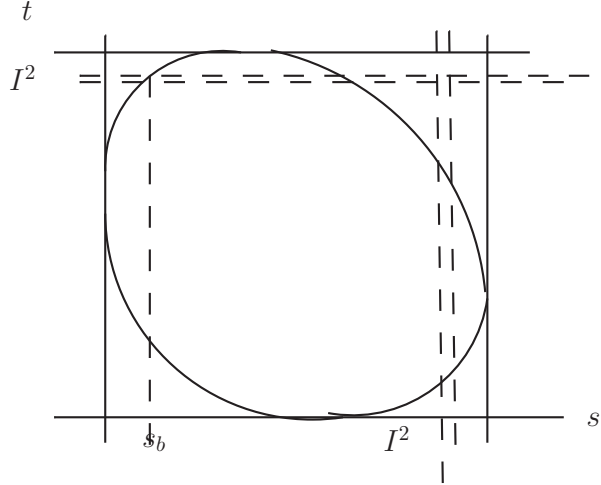} 
    \end{center} 
    \caption{Identical resonances in the $s$- and $t$-channels.}
    \end{figure} 
 
 We need to consider, instead, a case where the $s$ and $t$ channels, say, contain different interactions. An example of this type of triangle graph was calculated in \cite{me64}; see also \cite{anis}. The reaction considered was $\pi N \to \pi \pi N$, and $f(s, I^2, m^2)$ was the triangle shown in figure 27, where the
  \begin{figure}
    \begin{center} 
    \includegraphics[scale=.8]{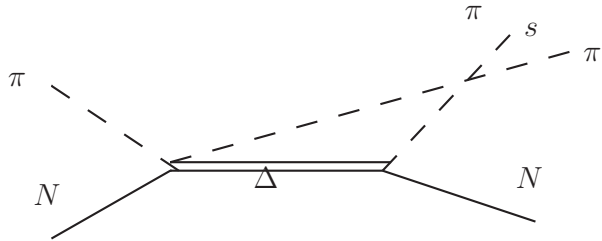} 
    \end{center} 
    \caption{A triangle calculated in \cite{me64} .}
    \end{figure} 
  intermediate resonance was the $\Delta(1232)$. All particles were treated as spinless, interacting in $s$-waves only. The triangle was calculated using a dispersion relation in $s$, just as in figure 19 for $f(s, \lambda^2, m^2)$: see figure 28. Both singularities $s_a$ and $s_b$ showed up clearly in
   \begin{figure}
     \begin{center} 
     \includegraphics[scale=.7]{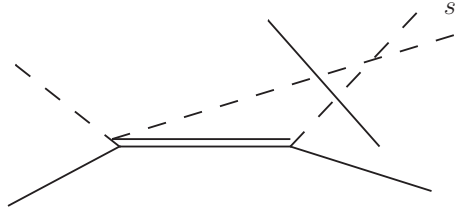} 
     \end{center} 
     \caption{Calculating figure 27 by dispersing in $s$.}
     \end{figure} 
   the integrand of $f(s, I^2, m^2)$, but only $s_b$ produced any effect in $|f(s, I^2, m^2)|^2$, and then only when $I^2$ was in the expected range
 \be 
 \frac{1}{2} (m^2+M^2 -2) \leq I^2 \leq (m-1)^2  \label{eq:Isqrange2}
 \ee
 which is the analogue of (\ref{eq:Isqrange}) for this  case (the pion mass is unity, and that of the nucleon is $M$). The sharpness of the $s_b$ effect depends sensitively on the width of the resonance $I$. For a realistic $\Delta$ width, no separate peak near $s=s_b$ was seen in $|f(s, I^2, m^2)|^2$, only a rise near threshold. For a width of order one tenth of the true width, a peak in the intensity near $s_b$ was present. In practice, since the width is also an overall factor in $T_{\rm sing}(s, m^2)$ of (\ref{eq: discD}), there will be a trade-off between the closeness of $s_b$ to the physical region and the magnitude of the effect. 
 
 In the 1960s and 1970s considerable effort went into trying to find a reaction in which a triangle singularity might be detectable. To my knowledge, no such effect was ever conclusively demonstrated in those days. More recently, however, the idea has been revived in various contexts. For example, Szczepaniak \cite{scz} considers diagrams of the type shown in figure 29, where his notation for the
  \begin{figure}
    \begin{center} 
    \includegraphics[scale=.7]{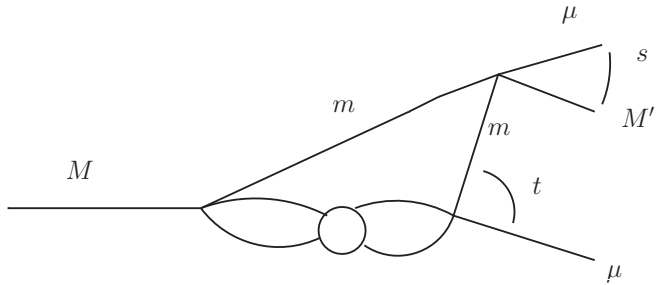} 
    \end{center} 
    \caption{The triangle considered in \cite{scz}.}
    \end{figure} 
  masses is used. He calculates two cases. In the first, ``$M$'' is the Y(4260) state, ``$m$'' is the average of the D and D$^*$ masses, ``$\mu$'' is a (massless) pion, and ``$M'$'' is the ${\rm J}/\psi$. For a $t$-channel resonance ${\rm D}^*_0$ at mass 2.4 GeV, he finds an enhancement near the $s$-threshold due to the $s_b$ singularity, close to the ${\rm Z}_{\rm c}$(3900) seen in the (${\rm J}/\psi \pi$) final state. All spins were neglected, and interactions were in $s$-waves. In a second case, Szczepaniak takes ``$M$'' to be the $\Upsilon$(5s, 1086), ``$m$'' to be the average of the B and B$^*$ masses, and ``$M'$'' to be the $\Upsilon(1s)$. A $t$-channel B$^*$ resonance at 5.698 GeV produces an enhancement near threshold in the $\Upsilon(1s) \pi$ channel, in the region of the observed ${\rm Z}_{\rm c}(10610)$ peak. 
 
 \subsubsection{ Enhancements in the three-body ($m^2$) channel} 
 
 So far we have concentrated on the possibility of a significant effect in the subenergy variable $s$, as $m^2$ varies. In cases where the final state allows different resonances in the $s$- and $t$-channels, rescatterings of the type shown in figure 30 will occur. Here $I^2_t$ and $I^2_s $ are the squared
  \begin{figure}
    \begin{center} 
    \includegraphics[scale=.7]{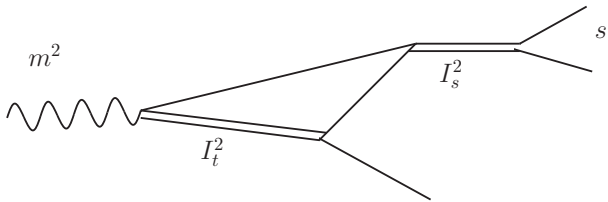} 
    \end{center} 
    \caption{Case of two different resonances in the $s$ and $t$ channels.}
    \end{figure} 
  masses of the resonances in the two channels. In such a case $f(s, I^2_t, m^2)$ will effectively be evaluated at $s=I^2_s$, and the rescattering amplitude $f(I_s^2, I^2_t, m^2)$ will exhibit a singularity in $m^2$ at $m^2=m_b^2$ say, near to the physical region in $m^2$ if the $I^2_s$ and $I^2_t$ bands cross on the ``magic'' upper left hand arc of the Dalitz plot (which only happens for one value of $m^2$): see figure 31.
   \begin{figure}
     \begin{center} 
     \includegraphics[scale=.7]{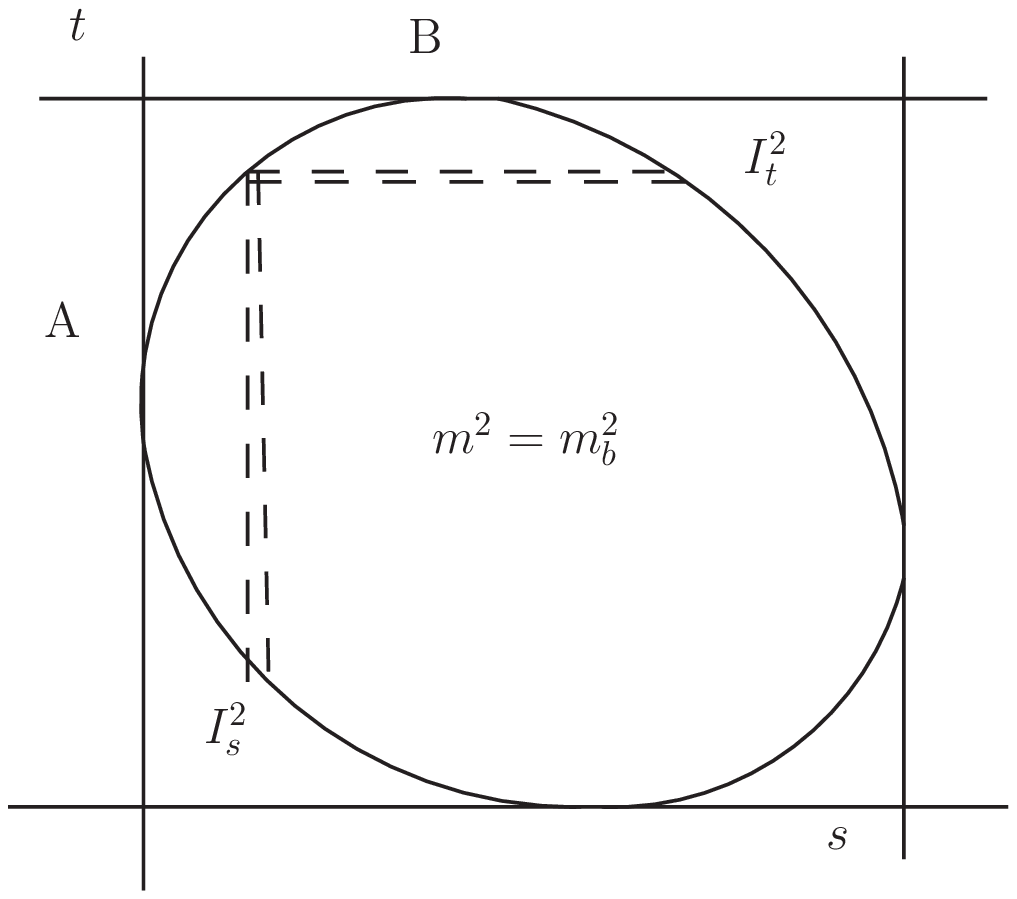} 
     \end{center} 
     \caption{The critical value $m^2=m_b^2$ such that the resonances in the $s$ and $t$ channels cross on the rescattering arc AB.}
     \end{figure}  
 
 One such process, considered in \cite{me64}, is shown in figure 32. In general, the $m^2$-enhancement 
  \begin{figure}
    \begin{center} 
    \includegraphics[scale=.7]{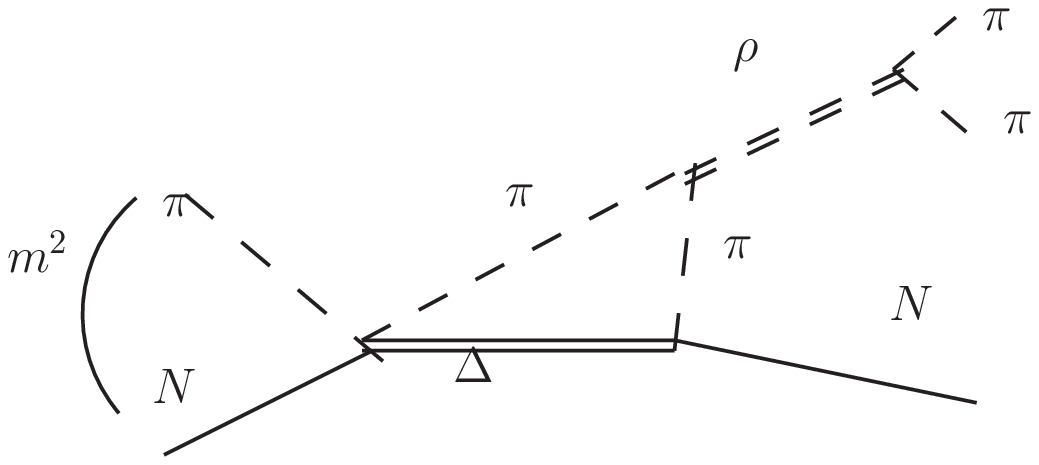} 
    \end{center} 
    \caption{The $m^2$-enhancement triangle considered in \cite{me64}.}
    \end{figure} 
 will occur near the $m^2$ threshold, in this case at $m^2 = (m_{\rm N} + m_\rho)^2$. There are several baryon resonances in this energy region, but their dynamical origin is different from the triangle singularity. 
 
 Recently it has been suggested \cite{MKS} that the ${\rm a}_1(1420)$ may be identified with such a triangle enhancement. Here the diagram is shown in figure 33. Spin and kinematic factors are included, and a peak in $|f|^2$, with a sharp phase motion relative to a reference wave, is found. The calculated effect (roughly $1\%$ in peak intensity)  is consistent with the data \cite{Com}. 
  \begin{figure}
    \begin{center} 
    \includegraphics[scale=.7]{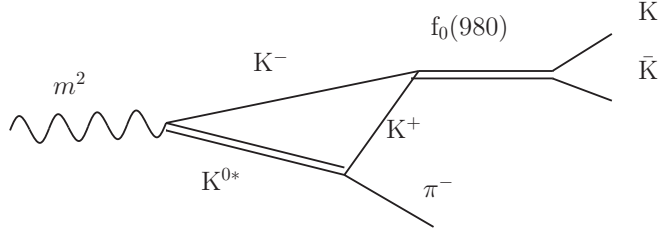} 
    \end{center} 
    \caption{The triangle graph identified with the ${\rm a}_1(1420)$ in  \cite{Com}.}
    \end{figure} 
 
 \subsubsection{ The observability of triangle singularities, and Watson's theorem}
 
 Consider a simple model in which there is a $t$-channel resonance in the amplitude $M_{13}(t)$, but not in the $s$-channel amplitude $M_{23}(s)$, and no interaction in the $u$-channel. Then including the first rescattering correction with a logarithmic singularity $s_b$, we have 
 \be 
 F(s,t,m^2) = C(m^2)\{M_{23}(s)[1 + T(s, m^2)] + M_{13}(t)\} \label{eq:Fobs} 
 \ee 
 where 
 \be 
 T(s, m^2) = \frac{1}{\pi} \int {\rm d} s' \frac{\rho(s')}{s'-s-{\rm i} \epsilon} \, B_{13}(s', m^2) \label{e q:Tobs} 
 \ee 
 and 
 \be 
 B_{13}(s', m^2) = \frac{1}{2} \int_{-1}^1 {\rm d} x_s \, M_{13}(t) = \frac{1}{2} \int_{-1}^1 {\rm d} x_s \frac{4 \gamma \sqrt{s_{\rm r}}}{I_t^2 - t(s', x_s)}. \label{eq:B13} 
 \ee 
 
 The discussion of Watson's theorem refers to a particular partial wave in a two-body channel. So consider the $l=0$ partial wave projection of $F(s, t, m^2)$ in the 2-3 cms. This is 
 \be 
 F_0(s, m^2) = \frac{1}{2} \int_{-1}^1 F(s, t, m^2) {\rm d} x_s = C(m^2) \{ M_{23}(s)[1+T(s, m^2)] + B_{13}(s, m^2)\}. \label{eq:F0}
 \ee 
 We now observe that the {\em same} function $B_{13}(s, m^2)$ appears in the projection (\ref{eq:F0}) and in the dispersion integral for $T(s, m^2)$. But whereas $B_{13}(s, m^2)$ has both singularities $s_a$ and $s_b$ near the physical region, as we saw in section 5.3.1, for $T(s, m^2)$  only the singularity at $s_b$ may be near the physical region, and then only for a range of $I_t^2$. Focusing then on the case in which $s_b$ is close to the physical region, we may ask: what is the net effect of having the $s_b$ singularity present in both $T(s, m^2)$ and $B_{13}(s, m^2)$? 
 
 This was the question raised and answered by Schmid \cite{Schmid}, and further discussed in \cite{claude2}. Using the identity (\ref{eq:dirac}), we can always write $T(s, m^2)$ as 
 \be 
 T(s, m^2) = {\rm i} \rho(s) B_{13}(s, m^2) + \frac{{\rm P.V.}}{\pi} \int {\rm d} s' \frac{\rho(s') B_{13}(s', m^2)}{s' - s}. 
 \label{eq:PV} 
 \ee
 The surprising fact is that it can be shown \cite{Schmid} \cite{claude2} that near the point $s=s_b$, when $s_b$ is near the physical region, the Principal Value integral in (\ref{eq:PV}) contributes equally with the $\delta$-function, so that 
 \be 
 T(s \approx s_b, m^2) \approx 2 {\rm i} \rho(s) B_{13}(s, m^2). 
 \label{scmidcond}
 \ee 
 Then $F_0(s, m^2)$ of (\ref{eq:F0}) becomes 
 \bea
 F_0(s, m^2) &\approx& C(m^2)\{ M_{23}(s) + B_{13}(s, m^2)[1+2 {\rm i} \rho(s) M_{23}(s)]\} \nonumber \\
 &=& C(m^2)[M_{23}(s) + B_{13}(s, m^2) {\rm e}^{2 {\rm i} \delta_{23}(s)}]. \label{eq:watsonresc}
 \eea 
 Thus the net effect of the nearby singularity $s_b$ in the rescattering correction to the projected amplitude $F_0$ is simply to modify the phase of the projection, $B_{13}(s, m^2)$, of the $t$-channel resonance. This was Schmid's result \cite{Schmid}, confirmed in \cite{claude2}. 
 
 Put differently, the intensity without the rescattering would be proportional to $|M_{23}(s) + B_{13}(s, m^2)|^2$, and with the rescattering to $|M_{23}(s) + {\rm e}^{2 {\rm i} \delta_{23}(s)} B_{13}(s, m^2)|^2$. While these may differ in magnitude, the presence of the rescattering singularity in $T(s, m^2)$ cannot be distinguished from its presence in $B_{13}(s, m^2)$. It would seem that the only surviving observable effect of the triangle singularity is the modification of the interference between 
 $M_{23}(s)$ and $B_{13}(s, m^2)$. Though a subtle effect, it may be relevant to experiments seeking to extract phase information from Dalitz plot interferences. 
 
 In concluding this section, we return to Watson's theorem. It is clear from (\ref{eq: discD}) that the phase of $F_0(s, m^2)$ is certainly not $\delta_{23}(s)$ for two reasons: first, the projection $B_{13}(s, m^2)$ is complex, and second so is the rescattering term $T(s, m^2)$.  
 
 \subsection{The single variable representation for $\Phi$ (1) }

Although, as we said, one could proceed on the basis of equations like (\ref{eq:Phidr}) as it stands, this involves a double integral on the RHS. It seems desirable to convert (\ref{eq:Phidr}) into a single variable integral equation, if possible. This will be more convenient for numerical work, and it also turns out to be much better suited for discussing general properties of the model - in particular the perhaps surprising fact that it can satisfy {\em three-body} unitarity as well. So we now turn to the single variable representation \cite{me65} for $\Phi$ (or $\phi$).

The single variable representation (SVR)  was first obtained in \cite{me65}, and we shall outline that derivation here. In section 5.5 we shall discuss an alternative, more general, derivation  given by Pasquier and Pasquier \cite{PP1}. 

In the present approach, the key step (due to Anisovich \cite{anis2}) exploits the fact that $\Phi(t, m^2)$ is   analytic in the $t$-plane cut along the real axis $t \geq 4$, so that we can write (always assuming convergence) 
\be 
\Phi(t,m^2) = \frac{1}{2 \pi {\rm i}} \int_C \frac{\Phi(\lambda^2, m^2)}{\lambda^2 - t} {\rm d} \lambda^2    \label{eq:Philam}
\ee 
where the  contour $C$ loops around the cut in a clockwise sense (c.f. figure 11). Then the integral term in (\ref{eq:Phidr}) becomes
\be 
2 \frac{M(s)}{2 \pi^2 {\rm i}} \int_4^\infty \frac{\rho(s')}{s'-s-{\rm i} \epsilon}\, \frac{1}{2} \int_{-1}^1 {\rm d} x_s \int_C \frac{\Phi(\lambda^2, m^2)}{\lambda^2 - t(s', x_s)} {\rm d} \lambda^2
\label{eq:tripint}
\ee 
which seems to have made matters worse. But we are going to invert the orders of integration in 
(\ref{eq:tripint}), after which things will look better. 

To do that, we need to be careful about the way the various singularities of the integrand are situated, with respect to the integration contours. A useful trick is to use a form of the third ingredient of $S$-matrix theory, namely crossing symmetry. In the present case, this will assert that our decay amplitude for $m \to 1+2+3$, with $m>3$, is the analytic continuation in $m$ of the $2 \to 2$ amplitude for $m+{\bar{1}} \to 2+3$ which is shown in figure 16. In practice, this means starting at a value $m < 3$, where the decay is not possible, doing the contour shuffling, and continuing the result
\begin{figure} 
\begin{center} 
\includegraphics{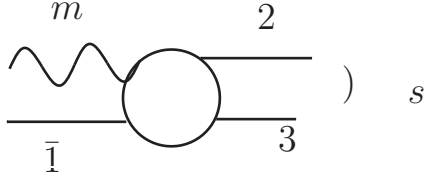} 
\end{center} 
\caption{The crossed process $m + {\bar{1}} \to 2 + 3$.}
\end{figure} 
 to a value $m >3$. After this manoevre, (\ref{eq:tripint}) becomes 
\be 
2 \frac{M(s)}{2 {\rm i}} \int_C \Phi(\lambda^2, m^2) f(s, \lambda^2, m^2){\rm d} \lambda^2    \label{eq:Phif}
\ee 
which indeed has $\Phi$ under a single integral, multiplied by the function 
\be 
f(s, \lambda^2, m^2) = \frac{1}{\pi^2} \int_4^\infty {\rm d} s' \frac{\rho(s')}{s' - s -{\rm i} \epsilon} \, \frac{1}{2} \int_{-1}^1 \frac{{\rm d} x_s}{\lambda^2 - t(s', x_s)}.  \label{eq:trigraph} 
\ee 
The function $f(s, \lambda^2, m^2)$ is again just  the triangle graph of figure 18 (up to conventional constants). 

We are now going to distort the contour $C$. To do this, we need to know about the singularities of $f(s, \lambda^2, m^2)$ as a function of $\lambda^2$. This is a rather technical matter, but the most important singularity is easy to understand. Looking at figure 18 along the $m^2$ direction, we can see that there is a threshold at $m^2 = (\lambda +1)^2$, which suggests that $f$ has a singularity at 
$\lambda^2 = (m-1)^2$.

We can verify the existence of the $\lambda^2 = (m-1)^2$  singularity directly from the representation (\ref{eq:trigraph}). We first rewrite the RHS of (\ref{eq:trigraph})  as 
\be
\frac{1}{\pi^2} \int_4^\infty \frac{{\rm d} s'} {s'-s-{\rm i} \epsilon}\, 
\frac{1}{\{[s' - (m-1)^2][s'-(m+1)^2]\}^{1/2}} \int_{t_-(s', m^2)}^{t_+(s', m^2)} \frac{{\rm d} t}{\lambda^2 - t} 
\label{eq:RHSf}
\ee
using (\ref{eq:tsx}) - (\ref{eq:qdef}). Here $t_{\pm}(s', m^2)$ are the phase space limits in $t$ for a given $s'$ and $m^2$:
\be 
t_\pm (s', m^2) = (3+m^2-s')/2 \pm 2p(s', m^2) q(s'). 
\label{eq:pslim}
\ee  
The rightmost integral in (\ref{eq:RHSf}) is simply 
\be 
 \left( \frac{\lambda^2 - t_-(s', m^2)}{\lambda^2 - t_+(s', m^2)} \right)   \label{eq:Log} 
\ee
which has singularities in $s'$ when $\lambda^2 = t_\pm (s', m^2)$, which is just the boundary of the Dalitz plot in the $s'-\lambda^2$ variables. So the singularities in $s'$ are at $s'_\pm(\lambda^2, m^2)$, which are the intersections of a fixed $\lambda^2$ line with the boundaries of the $s'-\lambda^2$ plot. The location of these singularities is what we need to understand. 
 
They can be visualised from figure 14, if we mentally replace $s$ by $\lambda^2$ and $t$ by $s'$. In particular, in the crossed region  
 $\lambda^2 \geq (m+1)^2$,  the intersections $s'_\pm(\lambda^2, m^2)$ are both negative, and do not interfere with the integration region $s' \geq4$ in (\ref{eq:RHSf}). Thus for large $\lambda^2 > (m+1)^2$ we run into no $\lambda^2$ singularities of $f$. 

Now consider reducing $\lambda^2$. At $\lambda^2 = (m+1)^2$, the points $s'_\pm(\lambda^2, m^2)$ coincide at the point $s'=1-m$, and then for $(m-1)^2 < \lambda^2 < (m+1)^2$ they become complex, one with a positive imaginary part and one with a negative imaginary part, as shown in figure 35. The two intersections meet again when $\lambda^2 = (m-1)^2$, at the point
\begin{figure} 
\begin{center} 
\includegraphics{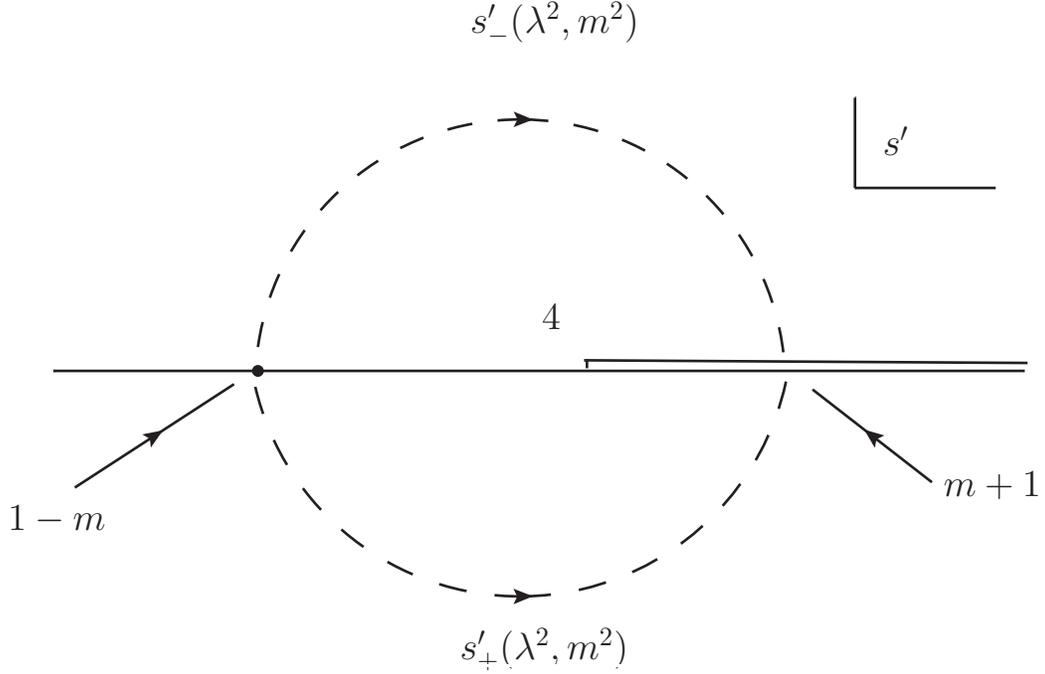} 
\end{center} 
\caption{The motion of the singularities $s'_{\pm}(\lambda^2, m^2)$ for $(m+1)^2 \leq \lambda^2 \leq (m-1)^2$.} 
\end{figure} 
$s' = m+1$, which for $m >3$ is 
beyond the start of the $s'$ integration in  $f$. One of $s'_\pm(\lambda^2, m^2)$ approaches $s'=m+1$ from  just above the contour, the other from just below. This means that the contour is ``pinched'', which is why there is a singularity of $f$ at $\lambda^2 = (m-1)^2$. 

Note, now, that for $m > 3$ the singularity at $\lambda^2 = (m-1)^2$ lies to the right of the point $\lambda^2=4$, so we need to know whether it lies above or below the $\lambda^2 \geq 4$ cut of $\Phi(\lambda^2, m^2)$. The answer is that the physical limit for this decay process is taken in the sense of $m^2 + {\rm i} \epsilon$. We therefore position the branch point at $\lambda^2=(m-1)^2$ above the $\lambda^2 \geq 4$ cut, as shown in figure 36. 
\begin{figure}
\begin{center}
\includegraphics{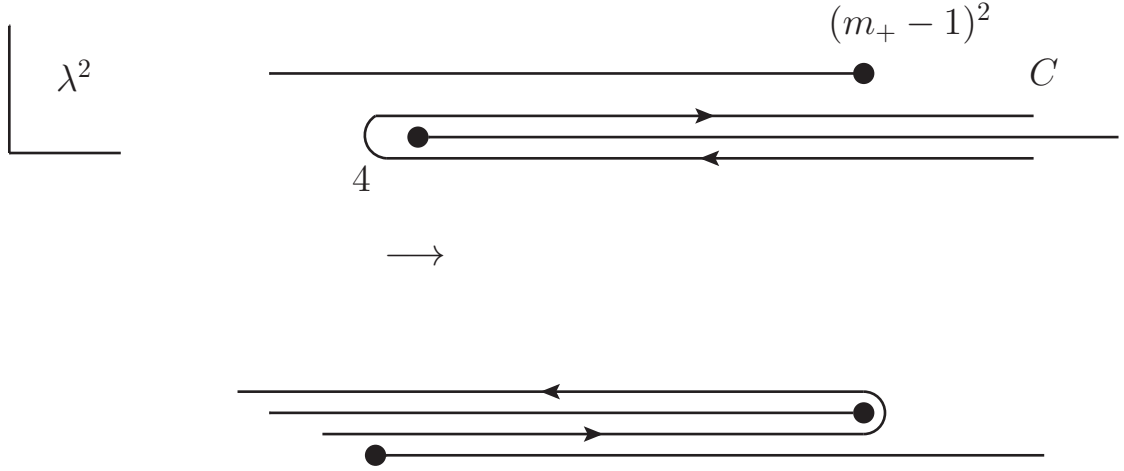}
\end{center}
\caption{The original contour $C$ in (\ref{eq:Philam}), and the distorted contour.}
\end{figure}

We now distort the contour $C$ so as to wrap around the $\lambda^2 \leq (m-1)^2$ cut as shown in figure 36. We end up, at this stage, with (\ref{eq:Phif}) replaced by 
\be 
2 M(s) \int_{-\infty}^{(m-1)^2} {\rm d} \lambda^2 \Phi(\lambda^2, m^2) \Delta_1( \lambda^2, m^2,s) 
\label{eq:corrDelta}
\ee 
where $\Delta_1( \lambda^2, m^2,s)$ is $1/2{\rm i}$ times the discontinuity of $f(s, \lambda^2, m^2)$ across the $\lambda^2 = (m-1)^2$ cut.

The discontinuity of $f$ across the $\lambda^2 = (m-1)^2$ cut can be calculated in various ways. Just as we guessed the existence on the $\lambda^2 = (m-1)^2$ singularity from inspection of figure 18, we can guess that the discontinuity across the associated cut will be found by cutting the graph as in figure 37. The result will then be proportional  to the
\begin{figure} 
\begin{center} 
\includegraphics{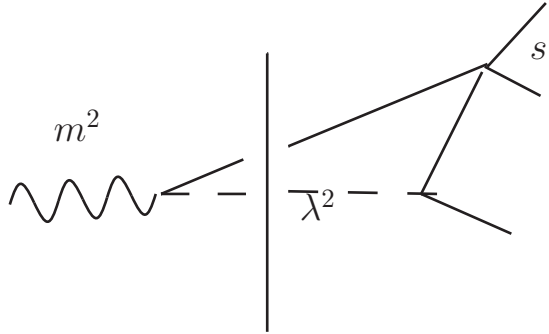} 
\end{center}
\caption{Graphical representation to get the discontinuity of $f(s, \lambda^2, m^2)$ across the $\lambda^2 =(m-1)^2$ cut.} 
\end{figure} 
  product of (a) the phase space factor for the intermediate on-shell state of one particle of unit mass and a second particle of mass $\sqrt{\lambda^2}$, with squared c.m.s. energy $m^2$, and (b) the $s$-wave projection, in the $m^2$-channel c.m.s., of the one particle exchange process shown in figure 38. In this exchange, the momenta are such that 
  \begin{figure} 
  \begin{center} 
  \includegraphics{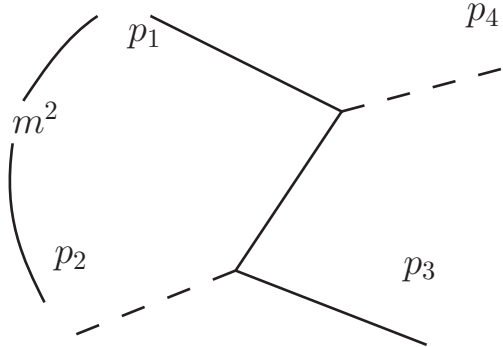} 
  \end{center} 
  \caption{One-particle exchange graph entering into the $\lambda^2$-discontinuity of 
  $f(s, \lambda^2, m^2)$.} 
  \end{figure} 
  \be 
p_1^2=p_3^2=1, \ \ \ p_2^2 = \lambda^2,\, p_4^2 = s, \ \ \ {\rm and} \ \ (p_1 +p_2)^2 = m^2.\label{eq:realexch}
\ee 
This is basically correct: from standard techniques of Feynman graph analysis \cite{me65} \cite{claude3},  the required discontinuity is calculated to be $2{\rm i} \Delta_1(s, \lambda^2, m^2)$ where 
\be
\Delta_1( \lambda^2, m^2,s) = \frac{1}{\pi k(s, m^2)} \ln \left( \frac{R-\sqrt{U}}{R+\sqrt{U}} \right) 
\label{eq:Delta1}
\ee 
where $k(s,m^2)$ is as in (\ref{eq:psmsq}), 
\be 
R(s, \lambda^2, m^2) = -m^4 +m^2(s+\lambda^2) + (\lambda^2-1)(s-1) \label{eq:R}
\ee 
and 
\be 
\sqrt{U(s, \lambda^2, m^2)}= k(\lambda^2, m^2) k(s, m^2). \label{eq:sqrtU}
\ee 
(A technical detail in parenthesis: the physical region for an external kinematic variable like $s$  is  in the sense of $s+{\rm i} \epsilon$, but for an internal variable like $\lambda^2$ it is in the sense $\lambda^2 - {\rm i} \epsilon$. This implies that the relevant discontinuity in $\lambda^2$ will actually be the difference  ``below the cut - above the cut'' \cite{claude3}. It is just this discontinuity that we need in figure 36.)

On the other hand, the $s$-wave projection in the $m^2$-c.m.s. of the exchange process of figure 38 is 
\be
\alpha( \lambda^2, m^2,s) =  \frac{1}{2} \int_{-1}^1 {\rm d} y \, \frac{1}{1-(p_1-p_4)^2}   \label{eq:alpha} 
\ee 
where $y$ is the cosine of the angle between ${\bm p_1}$ and ${\bm p_4}$ in the $m^2$-c.m.s. We obtain 
\be 
\alpha(\lambda^2, m^2, s) = \frac{m^2}{k(s, m^2) k(\lambda^2, m^2)} \ln \left( \frac{ R- \sqrt{U}}{R+ \sqrt{U}} \right).   \label{eq:alpha1} 
\ee 
Further, the two-particle phase space associated with the $p_1-p_2$ state is proportional to 
\be
\sigma(\lambda^2, m^2) = \frac{k(\lambda^2, m^2)}{m^2}. \label{eq:sigma} 
\ee 
So we see that, as expected, the function $\Delta_1$ is proportional to the product $\sigma(\lambda^2, m^2) \alpha( \lambda^2, m^2,s)$. 

$\Delta_1$ will figure prominently in what follows. It clearly ``belongs'' in the $m^2$ (i.e. 3-body) channel. Up to a kinematic factor, it is the projection of a  single-particle exchange graph, with the unusual feature that its pole occurs in the physical region. For this reason it is often called a ``real particle exchange (RPE)'' process - meaning that the propagator in (\ref{eq:alpha}) can vanish in the physical region.This singularity shows up in (\ref{eq:alpha1}), which is singular when $R^2=U$. This can  be written as $4m^2 \Gamma(s, \lambda^2, m^2) = 0$, where $\Gamma$ is our old friend the Kibble cubic of (\ref{eq:kibble}). So $\Delta_1$ has logarithmic singularities on the boundary of the $s-\lambda^2$ decay region ${\mathcal D}$. Inside this region, $\Delta_1$ develops an imaginary part of ${\rm i} \pi$. As a result, $\Phi$ will carry a phase which is additional to that of the two-body amplitude $M$, which is supplied in the isobar model. This additional phase is a direct consequence of RPE processes in the three-body problem.

\subsection{The single variable representation for $\Phi$ (2)}

The treatment of the $\lambda^2=(m-1)^2$ singularity of $f(s, \lambda^2, m^2)$ was relatively simple, but there are other singularities of $f$ at $\lambda^2=0$, in this equal mass case. These were studied in 
\cite{me65}, \cite{claude3} and \cite{claude4}. However, in more general cases involving non-zero  angular momentum states, and particles with spin, a simple Feynman graph interpretation is not available. Instead, as mentioned earlier, Pasquier and Pasquier \cite{PP1} showed how the single variable representation (SVR) can be derived directly by manipulating the double integral in (\ref{eq:Phidr}). 

In the Pasquier method, one begins by rewriting (\ref{eq:Phidr}) as 
\bea 
\Phi(s, m^2)&=&M(s) + 2 M(s) \frac{1}{\pi} \int_4^{(m-1)^2} \frac{ {\rm d} s' }{s'-s}\times  \nonumber \\ &\times&\frac{1}{\{[s'-(m-1)^2][s'-(m+1)^2]\}^{1/2}}\int_{t_-(s', m^2)}^{t_+(s', m^2)} \Phi(t) {\rm d}t 
\label{eq:Phiint}
\eea 
where $t_{\pm}(s', m^2)$ are the phase space limits in $t$ for a given $s'$ and $m^2$:
\be 
t_{\pm}(s', m^2)= (3+m^2-s')/2 \pm 2 p(s', m^2) q(s').
\ee
The method proceeds by casting the double integral in (\ref{eq:Phiint}) into the form of two contour integrals, and then inverting the order of the integrations via a series of contour deformations. The result is that (\ref{eq:Phiint}) can be transformed into the single variable integral equation 
\bea
\Phi(s, m^2) &=& M(s) + 2 M(s) \{ \int_{-\infty}^{(m-1)^2} \Delta( \lambda^2, m^2, s)\Phi(\lambda^2, m^2) {\rm d} \lambda^2 + \nonumber \\
&+& \int_{-\infty}^0 L( \lambda^2, m^2, s) \Phi(\lambda^2, m^2) {\rm d} \lambda^2\}, \label{eq:Phipasq}
\eea 
where 
\be 
\Delta(\lambda^2, m^2, s) = \frac{1}{\pi} \int_{s_-(\lambda^2, m^2)}^{s_+(\lambda^2, m^2)} \frac{{\rm d}s'}{s'-s} \frac{1}{\{[s'-(m-1)^2][s'-(m+1)^2]\}^{1/2}} \label{eq:Delta}
\ee 
and 
\be 
L(\lambda^2, m^2, s)=\frac{1}{\pi} \int_{-\infty}^{s_-(\lambda^2, m^2)} \frac{{\rm d}s'}{s'-s} 
\frac{1}{\{[s'-(m-1)^2][s'-(m+1)^2]\}^{1/2}}. \label{eq:L}
\ee
Here $s_{\pm}(\lambda^2, m^2)$ are the intersections of the phase space boundary curve $\Gamma(s, \lambda^2, m^2)=0$ with lines of fixed $\lambda^2$. They may be visualized from figure 14, redrawn in the variables $s, \lambda^2$ instead of $s, t$. In the expression for $L$, $s_-(\lambda^2, m^2)$ is the intersection with the boundary of region III. The integral in (\ref{eq:Delta})  can be evaluated analytically \cite{claude3} to show that $\Delta$ is in fact the same as $\Delta_1$. The expression for $L$ can also be evaluated \cite{me77} in terms of similar functions, but we do not give the formulae here. Actually, in sections 6 and 7  we shall give reasons for omitting the $\lambda^2 \leq 0$ contribution in (\ref{eq:Phipasq}).  In this case, the integral equation (\ref{eq:Phipasq}) can be conveniently represented in diagrammatic form as in figure 39. 
\begin{figure}
\begin{center}
\includegraphics[scale=.8]{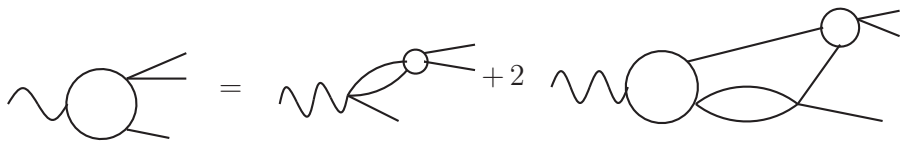}
\end{center} 
\caption{Diagrammatic representation of the integral equation for $\Phi(s, m^2)$ omitting the $\lambda^2 \leq 0$ contributions.}
\end{figure}

The Pasquier inversion was applied to the three-pion system by Pasquier and Pasquier \cite{PP2}, and to final states of the type ${\rm K} \pi {\rm N}$ (unequal masses, non-zero spin) by Brehm \cite{JJB1} and by Aitchison and Brehm \cite{AB1}, \cite{AB2}, \cite{AB3}. 

All the foregoing can be straightforwardly extended to various more complicated situations. Consider, for example, a model in which we have two pairs of final state particles interacting so as to form (different) isobars, but not the third pair. Then we write 
\be 
F(s,t, m^2) = M_1(s) \phi_1(s,m^2) + M_2(t) \phi_2(t, m^2)  \label{eq:F12} 
\ee 
where the discontinuity across the normal threshold in $\phi_1$ is 
\be 
{\rm disc} \ \phi_1 (s, m^2) = 2 {\rm i} \rho(s) \frac{1}{2} \int_{-1}^1 {\rm d} x_1 M_2(t) \phi_2(t, m^2)   \label{eq:discphi1} 
\ee 
and a  similar equation for ${\rm disc} \ \phi_2$. As before, we derive a single variable representation for $\phi_1$ and $\phi_2$ having the forms 
\bea 
\phi_1(s, m^2) &=& C_1(m^2) + \int K_{12} ( t', m^2, s) \phi_2( t', m^2) {\rm d} t'
\label{eq:phi1svr}  \\
\phi_2(t, m^2)& = & C_2(m^2) + \int K_{21} (s', m^2, t) \phi_1(s', m^2) {\rm d} s'    \label{eq:phi2svr} 
\eea 
where the inhomogeneous terms have now been chosen to reproduce the unmodified isobar terms in the two channels. 

However the equations (\ref{eq:phi1svr}) and (\ref{eq:phi2svr}) are not convenient for practical applications, because we would like to be able to calculate the corrections from a knowledge of the two-body interactions alone, independent of the ``fitting functions'' $C_1 $ and $C_2$. This is easy to arrange. If we iterate equations (\ref{eq:phi1svr}) and (\ref{eq:phi2svr}) we find that the equation for $F$ of (\ref{eq:F12})  can be rewritten as 
\bea 
F(s, t, m^2) &=& [M_1(s) \phi_{11}(s,m^2) + M_2(t) \phi_{21}(t, m^2) ] C_1(m^2) \nonumber \\
&+& [M_1(s) \phi_{12}(s, m^2) + M_2(t) \phi_{22} (t, m^2) ] C_2(m^2) \nonumber \\
&\equiv& C_1 \Psi_1(s, t, m^2) + C_2(m^2) \Psi_2(s,t,m^2). \label{eq:F12prod} 
\eea 
The quantities $\Psi_i$  describe rescattering starting in pair $i$, and ending in either pair $i$ or in pair $j$. The $\phi_{ij}$ describe corrections to be applied when an isobar is first produced in 
pair $j$ and rescatters finally to pair $i$. 
These functions satisfy equations of the following form:
\bea
\phi_{11}(s,m^2) &=& 1 + \int K_{12}(t', m^2, s) \phi_{21} (t', m^2) {\rm d} t'   \label{eq:phi11} \\
\phi_{21}(t, m^2) &=& \int K_{21}(s', m^2, t) \phi_{11} (s', m^2) {\rm d} s' \label{eq:phi21} \\
\phi_{12}(s, m^2) &=& \int K_{12} (t', m^2, s) \phi_{22} (t', m^2) {\rm d} t'   \label{eq:phi12} \\
\phi_{22}(t, m^2) &=& 1 + \int K_{21}(s', m^2, t) \phi_{12}(s', m^2) {\rm d} s'. \label{eq:phi22} 
\eea 

These coupled integral equations only depend on the two-body amplitudes $M_{i}$, and can be solved once and for all, leaving the $C_i(m^2)$ to be fitted to data.

It is clear that before any of this can be applied to experimental data, we must address the complications of  isospin and angular momentum. Both of these were introduced in a general way into 
this formalism by Pasquier and Pasquier \cite{PP2}. We now provide a brief introduction to these complications, and describe some calculations in physical systems.  

\section{Some Practical Examples }

We do not want to get too bogged down in the minutiae of 3-particle helicity states - which are contained in the references to be cited. We'll just give the general idea in the case of three spinless particles of unit mass.  

The {\bf first step} \cite{me77} is to generalise the expansion ({\ref{eq:Fcorr}) by writing 
\bea
&&F(s,t,u,m^2) = \sum_J(2J+1)\{ \nonumber \\
&&\; \; \; \sum_{\Lambda_1 l_1} {\mathcal{D}}^{J*}_{\Lambda_1 0}(\Omega_1) (2 l_1 +1) d^{l_1}_{\Lambda_10}(\theta_{12}) C_1^{J\Lambda_1 l_1}(m^2) \Phi^{J \Lambda_1 l_1} _1 (s, m^2) \nonumber \\
&&+ \sum_{\Lambda_2 l_2} {\mathcal{D}}^{J*}_{\Lambda_2 0} (\Omega_2) (2 l_2 +1) d^{l_2}_{\Lambda_2 0} (\theta_{23}) C_2^{J \Lambda_2 l_2}(m^2) \Phi^{J \Lambda_2 l_2}_2 (t, m^2) \nonumber \\
&&+ \sum_{\Lambda_3 l_3} {\mathcal{D}}^{J*}_{\Lambda_3 0} (\Omega_3) (2 l_3 +1) d^{l_3}_{\Lambda_3 0} (\theta_{31}) C_3^{J  \Lambda_3 l_3}(m^2) \Phi^{J \Lambda_3 l_3}_3 (u, m^2)\} \label{eq:Fangmom} 
\eea 
where $\Lambda_i$ are helicity labels (Cook and Lee \cite{cook}, Branson {\em et al.} \cite{BLT}, Berman and Jacob \cite{BJ}), $\theta_{ij}$ is the angle between the momenta of particles $i$ and $j$ in the $j-k$ c.m.s., and $\Omega_i$ specifies the orientation of ${\bm {p_i}}$ in the 3-body c.m.s. Also, $l_i$ is the pair partial wave and $J$ is the total angular momentum. 

We note that three different complete sets of states are employed in (\ref{eq:Fangmom}), so our basis is overcomplete. However, in each subenergy channel, only a finite number of partial waves will be retained, and one may regard the low partial waves in channels $t$ and $u$ as representing in some average sense the omitted high partial waves in channel $s$. In any case, the requirement of two-body (subenergy) unitarity will be imposed, and we shall see to what extent three-body unitarity can be satisfied also. 

The {\bf second step} \cite{me77} is to write down the subenergy unitarity relation, which takes the form 
\bea 
{\rm disc}_s \, \Phi_1^{J \Lambda_1 l_1} (s, m^2) &=& 2 {\rm i} \rho(s) M^{l_1 *}_{1}(s) \Phi^{J \Lambda_1 l_1}_1(s, m^2) \nonumber \\
&+& 2 {\rm i} \rho(s) M_1^{l_1 *} (s) \sum_{\Lambda_2 l_2} \frac{1}{2} \int_{-1}^1 C^J_{\Lambda_1 l_1 \Lambda_2 l_2} \Phi_2^{J \Lambda_2 l_2} (t, m^2) {\rm d} \cos \theta_{12} \nonumber \\
&+& {\rm similar} \ {\rm contribution} \ {\rm from} \ \Phi_3  \label{eq:discangmom} 
\eea 
where $C^J_{\Lambda_1 l_1 \Lambda_2 l_2}$ is proportional to the Wick \cite{wick} recoupling coefficient. If only the first term on the RHS of  (\ref{eq:discangmom}) were present, we would have the solution 
\be 
\Phi^{J \Lambda_1 l_1}_1(s, m^2) = G_1^{J \Lambda_1 l_1} (s, m^2) M_1^{l_1}(s) \label{eq:GM} 
\ee 
where $G$ does not have the $s \geq 4$ cut, but might have various kinematical factors (for example, centrifugal barrier factors as discussed by von Hippel and Quigg \cite{vHQ}). This is then the traditional isobar model. The rescattering corrections are incorporated by generalising (\ref{eq:GM}) to 
\be 
\Phi^{J \Lambda_1 l_1}_1(s, m^2) = G_1^{J \Lambda_1 l_1}(s, m^2) M_1^{l_1}(s)\phi_1^{J \Lambda_1 l_1}(s, m^2). \label{eq:phiangmom} 
\ee 

The {\bf third step} is to write a dispersion relation for $\phi_i$, and transform it into the single variable form. As an example, we write down the equation for the $J^P=1^-$ three-pion ($\omega$) channel (Aitchison and Golding \cite{me78} , Aitchison \cite{me77}):
\bea
\phi_\omega(s, m^2) &=& 1 + 2 \int_{-\infty}^{(m-1)^2} {\rm d} \lambda^2 \Delta_{1 \omega} (\lambda^2, m^2, s) M(\lambda^2) \phi_\omega(\lambda^2, m^2)   \nonumber \\
&+& {\rm contributions} \ {\rm from} \ \lambda^2 \leq 0 \label{eq:phiomega} 
\eea 
where 
\be 
\Delta_{1 \omega}(\lambda^2, m^2, s) = \frac{3}{2 k^2(s, m^2)}(\Gamma(s,t,u, m^2) \Delta_1(\lambda^2, m^2, s) - R k(\lambda^2, m^2)/2m^2) 
\label{eq:Delta1omega} 
\ee 
and $\Gamma$ is the Kibble cubic function $stu - (m^2-1)^2$, $\Delta_1$ is as given in 
(\ref{eq:Delta1}), and $R$ is given in (\ref{eq:R}). Isospin recoupling is included in (\ref{eq:phiomega}). 

This equation (including the $\lambda^2 \leq 0$ pieces) was studied in detail by Aitchison and Golding \cite{me78}. We parametrised $M(s)$ as 
\be 
M(s) = \left[a + bq^2 + c q^4 + \frac{2 q^3}{\pi \sqrt{(s)}} \ln \left( \frac{\sqrt{s} + \sqrt{s-4}}{\sqrt{s} - \sqrt{(s-4)}} \right)\right]^{-1}  \label{eq: Mrichard} 
\ee 
where the log has an imaginary part of $-\pi$ for $s>4$, and where as usual $s=4 + 4q^2$. Note that the $l=1$ partial wave requires the threshold factor $q^3$. The parameters $a$ and $b$ are equivalent to the mass and width parameters of a B-W amplitude, and were chosen to fit the physical $\rho$ meson in the first instance. We also explored other values of the mass and width parameters, as an exercise. The parameter $c$ controls the convergence properties of the integral - or in other words the importance of the far left hand region $\lambda^2 < 0$. Similar calculations were reported in Pasquier's thesis \cite{P}, but unfortunately remain unpublished. 

We found that with $m_\rho = (766 -\frac{1}{2} \, {\rm i}\,133)$ MeV it was possible to dynamically generate an $\omega$ resonance at the physical $m^2$ value, but this required a small value of the parameter $c$, resulting in a strong dependence on $\lambda^2 <0$ contributions. We regarded this as 
unphysical: these left hand contributions can be thought of as mimicking short-range effects (as opposed to the long-range single pion exchange processes associated with the rescatterings), which 
originate in $q {\bar{q}}$ dynamics, not $ 3 \pi$ dynamics. Nevertheless, it was interesting that the machinery could actually generate a 3-body resonance, and the calculated width was satisfactory (but 
presumably more or less fixed by the phase space). 

These considerations led us to try omitting the short range $\lambda^2 <0$ part, so that our equation now reads 
\be 
\phi_\omega(s, m^2) = 1 + 2 \int_0^{(m-1)^2} \Delta_{1 \omega} (\lambda^2, m^2, s) M(\lambda^2) \phi_\omega(\lambda^2, m^2).  \label{eq:phiomegatrunc} 
\ee 
This kernel function $\Delta_{1 \omega} $ is equal to the appropriate projection of the one-pion exchange graph in $\pi \rho \to \pi \rho$, up to multiplicative kinematic factors, which means that (\ref{eq:phiomegatrunc}) does include all long-range rescatterings. The parameter $c$ is now not needed for convergence, and is set to zero, while $a$ and $b$ are chosen to fit the $\rho$ mass and width values. With this truncated equation, we expect - and find - much less effect in the $m^2$ channel (no $\omega$ resonance), but pretty much the same result as far as the $s$-variation is concerned, which is dominated by the long-range rescatterings.\footnote{This provides another reason why we may reasonably truncate the $\lambda^2$-integration at $\lambda^2=0$: the contributions from $\lambda^2 \leq 0$ are sensibly constant in $s$, and so may be absorbed into the ``production vertex'' $C(m^2)$.} The logarithmic singularity $s_b$ mentioned earlier is visible. The deviation in the magnitude of $\phi_\omega$ away from unity was generally of the order of 20 - 30 \%. A phase of some $20^{\circ}$  could be generated at  $s$-values in the vicinity of the $\rho$ resonance. 

The $J^P = 1^+$ ($a_1$) wave was also investigated by Pasquier \cite{P} and by Parker \cite{ken}, using the full equations, but the results are only available in these authors' theses. Three channels were included: $\pi \rho$ in both $s$ and $d$ waves, and $\pi \epsilon$ in $l=1$,  where $\epsilon$ was taken to be a broad low-mass isoscalar $l=0$ $\pi-\pi$ state. Substantial $m^2$ variation was found, as well as significant rescattering from $\pi \rho$ to $\pi \epsilon$. The $3 \pi$ problem was taken up again by Brehm \cite{JJB2} \cite{JJB3}, who formulated and solved the  integral equations for the $J^P=0^-$ $\pi \rho$ wave, and the $1^+ \pi \rho$ and $\pi \epsilon$ channels. He found substantial $m^2$ dependence in the $1^+$ case, confirming the calculations of Pasquier \cite{P} and of Parker \cite{ken}. 

The formalism has also been applied to final state interactions in $\pi N \to \pi \pi N$ (Brehm \cite{JJB1}, Aitchison and Brehm \cite{AB1} \cite{AB2} \cite{AB3}). The spin of the nucleon is a significant technical complication. However, the same steps ``isobar-type expansion + subenergy unitarity + analyticity $\to$ single variable integral equations for the correction functions'' can be .followed through. 
The $J^P= {\frac{1}{2}}^+, {\frac{1}{2}}^-, {\frac{3}{2}}^+, {\frac{3}{2}}^-$ states were treated. All isobar states likely to be important for total energy $W \leq 1.5$ GeV were included, namely $N \pi$ isobars $S_{11}, S_{31}, P_{11}$ and $P_{33}$, and $\pi \pi$ isobars in $s$-wave $I=0, 2$, and the $p$-wave $I=1$.  The full integral equations were formulated, but only the first iterations (i.e. triangle graph contributions) were calculated, since experience had shown that the bulk of the subenergy variation (though not the $m^2$ variation) is well accounted for by the triangles. 

The main conclusions were as follows. First, none of the corrections vary rapidly with the subenergy variable. In addition, the shape of the subenergy variation changes very smoothly as $W$ varies. This implies that although there may be some observable corrections to the subenergy spectra, their presence will not significantly distort extracted $W$-channel resonance behaviour. Thus the non-unitary isobar model was to a large extent justified, at least as the data then stood. That is not to say, however, that with vastly more data, and with a focus on interferences on the Dalitz plot, such corrections can continue to be neglected.  

Secondly, and in this connection, a number of characteristic subenergy variations were found - none very large, to be sure, but possibly significant nowadays. One such variation exhibited strong curvature at the subenergy threshold, the real and imaginary parts crossing over each other. This behaviour was found in cases where all orbital angular momenta (both $l$ and $L$) were zero. A simple parametrisation of this pattern is provided by the   ``scattering length'' form 
\be 
\frac{1}{1-{\rm i} a q_i} \label{eq:sl}
 \ee 
 where $q_i$ is the magnitude of the pair momentum in their c.m.s. The real and imaginary parts of (\ref{eq:sl}) cross at $q_i = a^{-1}$. Typical values of $a$ were in the range 0.5 - 1 fm. The zero angular momentum cases were re-examined by Brehm\cite{JJB4}, who solved coupled integral equations of the type shown in (\ref{eq:phi11}) - (\ref{eq:phi22}) for the relevant amplitudes. The results for the isobar correction factors were quite well represented by (complex) scattering length parametrisations. 
 
 A second characteristic variation occurred in which there was curvature near the maximum of the kinematically allowed region in the subenergy variable $s_i$. Equivalently, this is the same as peaking in the variable  $p_i$, where 
 \be 
 p_i = \{ [W -(\sqrt{s_i} + m_i)^2][W-(\sqrt{s_i}-m_i)^2]\}/2W  \label{eq:pi} 
 \ee 
 is the magnitude of the momentum of the isobar in pair $i$, in the 3-body c.m.s., and $m_i$ is the mass of the remaining third particle. A simple parametrisation of this effect is the form 
 \be 
 \frac {1}{(1+(p_i R)^L)^{\frac{1}{2}}}. \label{eq:piR}
 \ee 
 We found that $R$ was typically of order 1 fm. Actually, just such a factor is frequently introduced into isobar model analyses (along with a threshold factor $(p_iR)^L$), as discussed by von Hippel and Quigg \cite{vHQ}. That such factors can have an impact on the subenergy spectrum was noted by Longacre \cite{long}. 
 
 These calculations were done a good many years ago, but were never to my knowledge ever combined with a revised isobar-model fit to data. However, the equations implementing subenergy unitarity and analyticity are in place, and it has recently been stated (Battaglieri {\em et al.} \cite{Bat}) that ``with the much larger data sets available today, this issue is certainly worth revisiting''. 
 
 At several points in the foregoing the reader will have noticed that, although the initial thrust of the procedure was very much focused on the two-body subenergy channels, the end result apparently had 
 relevance to the three-body channel as well. The kernel functions $\Delta_1$ are clearly three-body in nature, and the rescattering series generated by iterations of the single variable integral equations obviously contain three-body intermediate states. The question then arises: to what extent do these equations also incorporate three-body unitarity? Traditional (Faddeev-type) approaches to three-body f.s.i. would of course have a starting point which automatically satisfies three-body unitarity. So it is fair to ask whether the present treatment, based on the pair channels rather than the three-body channel, is capturing all the relevant physics. In fact, we'll now see that, rather surprisingly, our amplitudes do (or at least can) satisfy an appropriate form of three-body unitarity. 
 
 \section{Unitarity in the Three-body Channel} 
 
 We return to the simple model of (\ref{eq:Phipasq}), retaining only the $\Delta=\Delta_1$ kernel (the other terms  will not affect the following argument):
 \be
 \Phi(s, m^2) = M(s) + 2 M(s)\int_{-\infty}^{(m-1)^2} {\rm d} \lambda^2 \Delta_1( \lambda^2, m^2,s)  \Phi(\lambda^2, m^2).
 \label{eq:PhiSVR2} 
 \ee 
  We are now interested in the $m^2$ behaviour of $F$ as given by (\ref{eq:Fcorr}). We shall take $C(m^2)$, the isobar production amplitude, to have no singularities in $m^2$, so that whatever three-body structure there is, is in $\Phi(s, m^2)$. 
 
 We first verify that $\Phi(s, m^2)$ has a singularity at the three-particle threshold $m^2=9$. Note that for $m^2 > 9$ the upper limit of the integration in (\ref{eq:PhiSVR2}) will lie to the right of the threshold $\lambda^2=4$, where $\Phi(\lambda^2,m^2)$ has a branch point and associated cut. The physical limit  is via the $m^2 + {\rm i} \epsilon$ prescription, and  in that case the $\lambda^2$ integration contour will lie above the $\lambda^2 \geq 4$ cut. If, instead, we give $m^2$ a negative imaginary part, $m^2 - {\rm i} \epsilon$, the $\lambda^2$ integration contour will lie below the cut, and the result will be different. Thus the function must have a branch point at $m^2 = 9$, with a discontinuity $\Phi(s_+, m^2_+) - \Phi(s_+, m^2_-)$, which we shall now calculate (we are taking the physical limit for $s$). 
 As usual, this discontinuity will be directly related to unitarity in the three-body channel. We shall only aim to give the flavour of the analysis: a more careful discussion is contained in Aitchison and Pasquier \cite{AP}, and an even more careful one in Pasquier and Pasquier \cite{PP1}. 
 
 From (\ref{eq:PhiSVR2}) it follows that, for $m^2 \geq 9$, 
 \bea 
 \Phi(s_+, m^2_+) - \Phi(s_+, m^2_-) &=& 
 2 M(s_+)  \{ \int_{\Gamma_+}^{(m_+-1)^2} {\rm d} \lambda^2 \Delta_1(\lambda^2, m^2, s_+) \Phi(\lambda^2, m^2_+) \nonumber \\
  && - \int_{\Gamma_-}^{(m_- -1)^2} \Delta_1(\lambda^2, m^2, s_+)\Phi(\lambda^2, m^2_-)\}  \label{eq:discmsqPhi1}
 \eea
 where $\Gamma_\pm$ are the two integration contours lying above and below the $\lambda^2 \geq 4$ cut, as shown in figure 40. We have omitted the + or - labels on the $\lambda^2$ and
 \begin{figure} 
 \begin{center} 
 \includegraphics{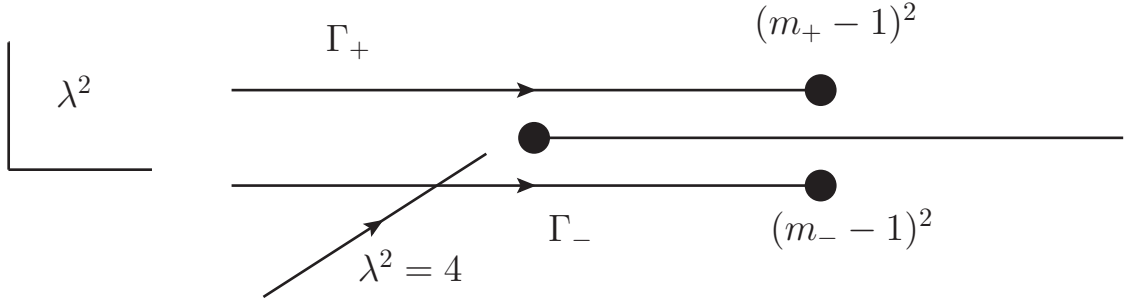} 
 \end{center} 
 \caption{The contours $\Gamma_{-}$ and $\Gamma_+$ in (\ref{eq:discmsqPhi1}).} 
 \end{figure} 
  $m^2 $ arguments of $\Delta_1$, since a more careful study shows that the singularities of $\Delta_1$ lie on the same (lower) side of both $\Gamma_+$ and $\Gamma_-$. However, we must use $\Phi(\lambda^2_+, m^2_+)$ on $\Gamma_+$, and $\Phi(\lambda^2_-, m^2_-)$ on $\Gamma_-$. Thus the RHS of (\ref{eq:discmsqPhi1}) is 
 \bea 2 M(s_+)  \int_{-\infty}^{(m-1)^2} {\rm d} \lambda^2 \Delta_1(\lambda^2, m^2, s_+) [\Phi(\lambda^2_+, m^2_+) - \Phi(\lambda^2_-, m^2_-)] \nonumber \\
 = 2 M(s_+) \{\int_{- \infty}^{(m-1)^2} {\rm d} \lambda^2 \Delta_1(\lambda^2, m^2, s_+) \{ [\Phi(\lambda^2_+, m^2_+) - \Phi(\lambda^2_+, m^2_-)] && \nonumber \\
  + [\Phi(\lambda^2_+, m^2_-)-\Phi(\lambda^2_-, m^2_-)]\}.&& \label{eq:discm1}
 \eea
 It follows that 
 \bea 
 {\rm disc}_{m^2=9} \Phi(s_+, m^2) = 2 M(s_+) \int_{- \infty}^{(m-1)^2} {\rm d} \lambda^2 \Delta_1(\lambda^2, m^2, s_+){\rm disc}_{m^2=9} \Phi(\lambda^2_+, m^2)&& \nonumber \\
  + 2 M(s_+) \int_{-\infty}^{(m-1)^2} {\rm d} \lambda^2 \Delta_1(\lambda^2, m^2, s_+) {\rm disc}_{\lambda^2=4} \Phi(\lambda^2, m^2_-)\ \ \ \ \  && \label{eq:discm2}
 \eea
 where ${\rm disc}_{\lambda^2 =4}\Phi(\lambda^2, m^2_-)$ means the discontinuity in $\Phi$ across the $\lambda^2 \geq 4$ cut, with the prescription $m^2 - {\rm i} \epsilon$. This latter is just the discontinuity given by the subenergy unitarity relation, but continued round from  $m^2+{\rm i} \epsilon$ to $m^2 - {\rm i} \epsilon$, namely (c.f. (\ref{eq:subendisc}))
 \be 
 {\rm disc}_{\lambda^2 =4}\Phi(\lambda^2, m_-^2) = 
 2 {\rm i} \rho(\lambda^2_+) M(\lambda^2_+) F^0(\lambda^2_-, m^2_-)\theta(\lambda^2-4)  \label{eq:discsubPhi} 
 \ee 
 where as before $F(\lambda^2, t, m^2) = \Phi(\lambda^2, m^2) + \Phi(t, m^2) + \Phi(u, m^2)$,  $u=3+m^2-\lambda^2-t$, and $F^0$ is the $s$-wave projection in the $\lambda^2$-channel c.m.s. of $F$:
 \be 
 F^0(\lambda^2, m^2) = \frac{1}{2} \int_{-1}^1 F(\lambda^2, t, m^2) {\rm d} x_{\lambda^2}.  \label{eq:Fproj} 
 \ee  
 
 So we arrive at an   integral equation for ${\rm disc}_{m^2=9}\Phi(s_+, m^2)$:
 \bea 
{\rm disc}_{m^2=9}\Phi(s_+, m^2)  =\ \ \ \ \ \ \ \ \ \ \ \ \ \ \ \ \ \ \ \ \  &&\nonumber \\
  2 M(s_+) \int_4^{(m-1)^2} {\rm d} \lambda^2 \Delta_1(\lambda^2, m^2, s_+) 2 {\rm i} \rho(\lambda^2_+) M(\lambda^2_+) F^0(\lambda^2_-, m^2_-)  && \nonumber \\
  + 2 M(s_+) \int_{-\infty}^{(m-1)^2} {\rm d} \lambda^2 \Delta_1(\lambda^2, m^2, s_+){\rm disc}_{m^2=9} \Phi(\lambda^2, m^2). \label{eq:inteqdisc} 
  \eea
  
 The first iteration of this equation is just the inhomogeneous term, which we write as  
 \bea
 {\rm disc}_{m^2=9}\Phi(s_+, m^2) = \hspace{1in} \nonumber \\
 2 \int_4^{(m-1)^2} {\rm d} \lambda^2 2 {\rm i} \rho(\lambda^2_+) \sigma(\lambda^2_+, m^2_+) F^0(\lambda^2_-, m^2_-) \Psi^{(1)}(\lambda^2_+, m^2_+, s_+) \label{eq:discmit1}
 \eea 
 where 
 \be 
 \Psi^{(1)}(\lambda^2, m^2, s) = M(\lambda^2) \frac{\Delta_1(\lambda^2, m^2, s)}{\sigma(\lambda^2, m^2)} M(s) \equiv Z(\lambda^2, m^2, s) \label{eq:Psi1}
\ee
and $\sigma$ was introduced in (\ref{eq:sigma}). Now we saw in section 5.4.1 that the quantity $\Delta_1/\sigma$ is proportional to the $s$-wave projection, in the three-body c.m.s., of the one-particle exchange process ``$\lambda^2 + 1 \to 1 +s$'' (see figure 38), and so $Z$ represents just the $s$-wave projection of figure 41, which is clearly the first term in a 
\begin{figure} 
\begin{center} 
\includegraphics{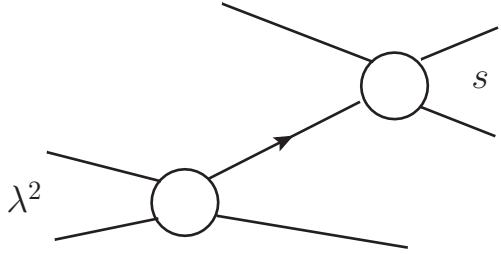} 
\end{center} 
\caption{Single-particle exchange diagram in the 3 $\to$ 3 amplitude.}
\end{figure}
three particle $\to$ three-particle scattering amplitude. Furthermore, the factor $\rho(\lambda^2) \sigma(\lambda^2, m^2)$ represents the expected phase space factor for the effective three-particle phase space (up to conventional constants). We recall from (\ref{eq:3ps}) that this phase space factor is proportional to 
\be 
\frac{{\rm d} \lambda^2 {\rm d} t} { m^2}   \label{eq:3bdyps}
\ee 
which, using (\ref{eq:tsx}) with $s$ replaced by $\lambda^2$ can be written as 
\be 
\frac{{\rm d} \lambda^2}{m^2} 2 p(\lambda^2, m^2) q(\lambda^2) {\rm d} x_{\lambda^2} = 
\sigma(\lambda^2, m^2) \rho(\lambda^2) {\rm d} \lambda^2 \, {\rm d} x_{\lambda^2}. 
\label{eq:3bdyps2}
\ee 
This is just the phase space appearing in (\ref{eq:discmit1}). The additional 2 in (\ref{eq:discmit1}) arises from the identical channels. 

Carrying out the iteration of (\ref{eq:inteqdisc}) to all orders, we find the result 
\bea 
{\rm disc}_{m^2=9}\Phi(s_+, m^2) = \hspace{1in} \nonumber \\
4{\rm i}  \int_4^{(m-1)^2} {\rm d} \lambda^2 F^0(\lambda^2_-, m^2_-) \rho(\lambda^2_+,) \sigma(\lambda^2_+, m^2_+) \Psi(\lambda^2_+, m^2_+, s_+)&& \label{eq:discmPhi3} 
\eea
where 
\be 
\Psi(\lambda^2, m^2, s) = Z(\lambda^2, m^2, s) + 2 M(s) \int_{-\infty}^{(m-1)^2}{\rm d} \mu^2 \Delta_1(\mu^2, m^2, s) \Psi(\lambda^2, m^2, \mu^2). \label{eq:Psi} 
\ee

Equation (\ref{eq:discmPhi3}) is a special case of the general discontinuity formula across a three-body cut, as given by Hwa \cite{hwa} and Fleming \cite{flem}. These discontinuities were in turn derived from three-body unitarity relations, together with subenergy unitarity relations. So it appears that in our approach, a combination of two-body unitarity, analyticity, and crossing have generated three-body unitarity in the ``pair interactions only'' approximation, and have also generated self-consistently a three particle to three-particle scattering amplitude. Naturally our amplitudes contain no three-body forces. Nevertheless, once having made that assumption, it is a viable option to work in the two-body channels, rather than in the generally more difficult three-body one, without sacrificing three-body unitarity. 

The amplitude $\Psi$ has a further interesting interpretation. By considering the iterative solution of (\ref{eq:PhiSVR2}), we find that $\Phi(s, m^2)$ can be written as 
\be 
\Phi(s, m^2) = M(s) + 2 M(s) \int_{-\infty}^{(m-1)^2} {\rm d} \lambda^2 M(\lambda^2) \psi(\lambda^2, m^2, s)  \label{eq:Phires} 
\ee 
where 
\be 
\psi(\lambda^2, m^2, s) = \Delta_1(\lambda^2, m^2, s) + 2 \int_{-\infty}^{(m-1)^2} {\rm d} \mu^2 \Delta_1(\mu^2, m^2, s) M(\mu^2) \psi(\lambda^2, m^2, \mu^2).  \label{eq:psi} 
\ee
 Equation ({\ref{eq:Phires}) shows that the amplitude $\psi$ is the resolvent of the integral equation for $\Phi$, where 
 \be 
 \psi(\lambda^2, m^2, s) = \sigma(\lambda^2, m^2) \Psi(\lambda^2, m^2, s) /[M(\lambda^2) M(s)]. 
 \label{eq:psiPsi}
 \ee  
 
 We need to add one important caveat concerning 3-body unitarity in this approach. While our amplitudes do satisfy the correct discontinuity relations, this only translates into a true 3-body unitarity relation if the $3 \to 3$ amplitude is symmetric, as pointed out by Pasquier and Pasquier \cite{PP1} \cite{PP2} - or at least can be made so by multiplying by a suitable kinematic factor. This is the case for the kernel function $\Delta_1$, but it is not the case for the kernels associated with the $\lambda^2 \leq 0$ part of the integral equations, which we have hidden from the reader. One can, indeed, symmetrise all the kernels by hand, but then the singularity structure of the amplitude  
 $\Phi$ will be changed. In fact, $\Phi$ will develop extraneous left-hand cuts in $s$ which are unphysical. The most attractive option is to simply truncate the $\lambda^2$ integral at $\lambda^2=0$, retaining only the $\Delta_1$ kernel - a strategy we already suggested in section 6.  In this case, of course, one is  again introducing new singularities, but they lie in the region $s \geq (m+1)^2 $, well into the inelastic region \cite{PP1}. One way of seeing that this is likely to be the case is to consider the particular rescattering graph of figure 42. If we want to reconstruct this process exactly
 \begin{figure} 
 \begin{center} 
 \includegraphics{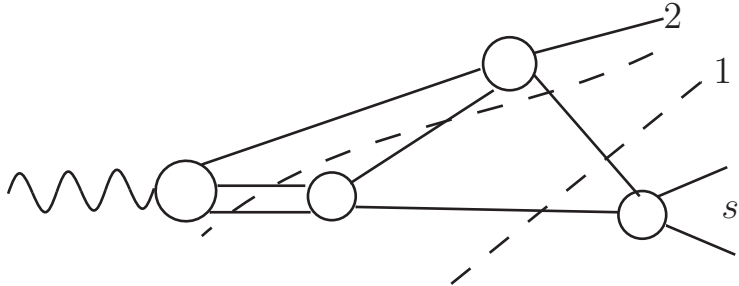}
 \end{center} 
 \caption{Rescattering amplitude.}
 \end{figure}
  by dispersing in the variable $s$, we need to include the discontinuities across not only the 2-body cut labelled 1, but also across the 4-body cut labelled 2. In the 3$\pi$ case in the $\omega$ and $a_1$ channels, satisfactorily symmetric kernels are found (Aitchison \cite{me77}, Pasquier and Pasquier \cite{PP2}). The same is true for the corresponding kernels in the meson + meson + nucleon case as was shown in Aitchison and Brehm \cite{AB1}, where their connection with the elementary one-particle exchange processes is also established. 
  
  The upshot of this discussion (see also section 6 and \cite{PP1}) is that the best practical form of the SVR for $\Phi$ is likely to be the {\em truncated} one in which the $\lambda^2 \leq 0$ contributions to the kernel are omitted. We remind the reader that in this case the equation has the diagrammatic representation of figure 39, clearly exhibiting the RPE graph in the kernel. 
 
 Although it is clearly desirable to have amplitudes which do satisfy three-body unitarity, in practice this constraint is usually implemented in phenomenological applications as a ``quasi two-body'' unitarity constraint, applied to the quasi two-body system consisting of an isobar and a third particle. Indeed, that was the procedure alluded to in section 3, in final states such as $\pi K^*, \rho K$, etc. In the sorts of systems we are here interested in, namely those  in which strong isobars are formed, the particle + resonance threshold is much more significant physically than the (uncorrelated) three-body threshold. But we should understand how this ``woolly'' quasi two-body unitarity is justified, at least in our model, and how it relates to the three-body structure in the $\Phi$ functions. This will be our last application of the single variable representation for $\Phi$. Such ``woolly cut'' discontinuity relations were discussed by many authors (Nauenberg and Pais \cite{NP}, Baz \cite{baz}, Ball, Frazer and Nauenberg \cite{bfn}, Zwanziger \cite{Zw} and  Frazer and Hendry \cite{fh}).   
 
 \section{Particle-Resonance Scattering} 
 
 The reduction of the three-body scattering problem to an effective two-body one by invoking the isobar model was first introduced by Mandelstam {\em et al.} \cite{MPPS}. In this approach, we first define what we mean by a ``particle + resonance'' amplitude. In the case of $\Phi(s, m^2)$, for example, where $\Phi(s, m^2) = M(s) \phi(s, m^2)$, we shall take $M(s)$ to have a resonance pole at $s=s_{\rm C}$ on the second sheet reached from the upper side of the $s \geq 4$ cut, as usual; and we also know that $\phi(s, m^2)$ has the same $s \geq 4$ cut but no resonance pole. Let us denote a value of $s$ on the second sheet of that cut by $s^{{\rm II}}$ in Zwanziger's notation. Then  
 $M(s^{\rm II})$ may be parametrised near the resonance as 
 \be 
 M(s^{{\rm II}}) \sim g^2/(s_{\rm C} - s^{{\rm II}})  \label{eq:Mres} 
 \ee 
 where $s_{\rm C}$ is complex with a negative imaginary part. We define our particle resonance amplitude $\phi(s_{\rm C}, m^2) $ as the residue of the pole in $\Phi(s, m^2)$ at $s^{\rm II}=s_{\rm C}$, divided by $g^2$:
 \be 
 \phi(s_{\rm C}, m^2) = \lim_{s^{\rm II}  \to s_{\rm C}} \frac{(s_{\rm C}- s^{\rm II})}{g^2} \Phi(s^{{\rm II}}, m^2). \label{eq:presphi} 
 \ee
We represent $\phi(s_{\rm C}, m^2)$ by figure 43. In a similar way, from our $3 \to 3$
\begin{figure} 
\begin{center} 
\includegraphics{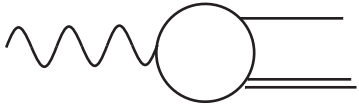} 
\end{center} 
\caption{Production amplitude for particle + resonance.} 
\end{figure}
 amplitude $\Psi(\lambda^2, m^2, s)$ (see (\ref{eq:Psi})) 
 we can define 
 \be 
 \psi(s_{\rm C}, m^2, s_{\rm C}) = \lim_{s^{\rm II} \to s_{\rm C}} \lim_{\lambda^{2{\rm II}} \to s_{\rm C}} 
 \frac{(s^{\rm II} - s_{\rm C})}{g^2} \frac{(\lambda^{2 \,{\rm II}} - s_{\rm C})}{g^2} 
 \sigma(\lambda^2, m^2) \Psi(\lambda^{2{\rm II}}, m^2, s^{{\rm II}}). 
 \label{eq:psires}
 \ee 
 
 We now turn to the analysis of the singularities of $\phi(s_{\rm C}, m^2)$ - in particular, we want to exhibit the fact that it has a branch point at $m^2 = (\sqrt{s_{\rm C}}+1)^2$, which is the threshold for the production of the particle + resonance state. But we must be precise about what sheet this branch point is on. 
 
 We write out our equation for $\Phi(s, m^2)$ once more, in the form 
 \be 
 \Phi(s, m^2) = M(s) + 2 M(s) \int^{(m-1)^2} {\rm d} \lambda^2 \Delta_1(\lambda^2, m^2, s) M(\lambda^2) \phi(\lambda^2, m^2) \label{eq:Phiagain} 
 \ee 
 where the integration contour in the $m^2 + {\rm i} \epsilon$ limit lies above the $\lambda^2 \geq 4$ cut of $\phi$ when $ m^2 > 9$. Consider now an analytic continuation in $m^2$, starting with $m^2 > 9$ and a small positive imaginary part, and then going down into the lower half plane, through the real axis. The $\lambda^2$ integration upper limit  $\lambda^2 = (m-1)^2$ will descend into the second sheet of $M(\lambda^2)$ and $\phi(\lambda^2, m^2)$, which now become $M(\lambda^{2\,{\rm II}})$ and $\phi(\lambda^{2 \, {\rm II}}, m^2)$. But $M(\lambda^{2 \, {\rm II}})$ has the pole at $s=s_{\rm C}$ on this  sheet, and it follows that there will be a singularity of $\Phi(s, m^2)$ in the variable $m^2$ when the end point at $\lambda^2 = (m-1)^2$ hits the pole at $\lambda^2 = s_{\rm C}$. This occurs when $m^2 = (\sqrt{s_{\rm C}} + 1)^2$, the particle + resonance threshold (we will see later that it is indeed a square root branch point). We can draw an associated cut as in figure 44. 
 \begin{figure} 
 \begin{center} 
 \includegraphics[scale=.7]{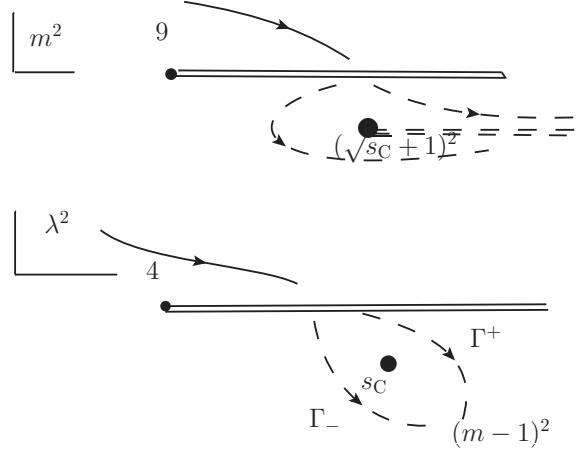}
 \end{center}
 \caption{Calculating the discontinuity across the particle + resonance cut.}
 \end{figure}
 
 The discontinuity across this cut is calculated, as in the case of the one across the $m^2 \geq 9$ cut, by considering the difference between an $m^2$-continuation to a point just above the cut (so that the $\lambda^2$ contour $\Gamma_+$ leaves the pole on its right hand side - see figure 44), and an $m^2$ continuation to a point just below the cut (so the contour $\Gamma_-$ leaves the pole on its left hand side). This difference is 
 \bea 
 \Phi(s_+, m^2_+) - \Phi(s_+, m^2_-) = 2 M(s_+) \times  \hspace{2in} && \nonumber \\
  \int^{(m-1)^2} {\rm d} \lambda^2 \Delta_1(\lambda^2, m^2, s_+) \{ M(\lambda^{2 \, {\rm II}}_+) 
 \phi(\lambda^{2 \, {\rm II}} , m^2_+) - M(\lambda^{2 \, {\rm II}}_-) \phi(\lambda^{2 \, {\rm II}}, m^2_-) \} 
 \label{eq: woolly} 
 \eea 
 where now $\lambda^{2 \, {\rm II}}_+ \, ( \lambda^{2 \, {\rm II}}_-)$ means that the pole in $M(\lambda^{2 \, {\rm II}})$ at $\lambda^2 = s_{\rm C}$ is always to the right (left) of the oriented contours $\Gamma_+$ and $\Gamma_-$, and where $m^2_+ \,(m^2_-)$ means that $m^2$ is above (below) the branch cut starting at $m^2 = (\sqrt{s_{\rm C}} +1)^2$.  The $\pm$ on the $\lambda^2$ argument of $\phi$  is irrelevant since it lacks the pole at   $\lambda^{2\,{\rm II}}=s_{\rm C}$.
  The difference in curly brackets is 
 \be 
 M(\lambda^{2 \, {\rm II}}_+) [\phi(\lambda^{2 \, {\rm II}}, m^2_+) - \phi(\lambda^{2 \, {\rm II}}, m^2_-)] + \phi(\lambda^{2\,{\rm II}}, m^2_-)[M(\lambda^{2 \, {\rm II}}_+) - M(\lambda^{2 \, {\rm II}}_-)]. \label{eq:woolydisc}
 \ee 
The difference $M(\lambda^{2 \, {\rm II}}_+) - M(\lambda^{2 \, {\rm II}}_-)$ is $ 2 \pi {\rm i} $ times the discontinuity on a full circuit around the pole, which is $2 \pi {\rm i} g^2 \delta(\lambda^{2 \, {\rm II}} - s_{\rm C})$. Hence (\ref{eq:woolydisc}) 
 becomes 
 \bea 
 [\Phi(s_+, m^2_+)-\Phi(s_+, m^2_-)] = 2 M(s_+) 2 \pi {\rm i} g^2 \phi(s_{\rm C}, m^2_-) \Delta_1(s_{\rm C} , m^2, s_+) && \nonumber \\
 + 2 M(s_+) \int^{(m-1)^2} {\rm d} \lambda^2 \Delta_1(\lambda^2, m^2, s_+)[\Phi(\lambda^{2 \, {\rm II}}_+, m^2_+) - \Phi(\lambda^{2 \, {\rm II}}_+, m^2_-)]. 
 \label{eq:woolydiscinteq}
 \eea 
  
  Once again, this is an integral equation for the required discontinuity. The first iteration is 
  \be 
  [\Phi(s_+, m^2_+) - \Phi(s_+, m^2_-)]^{(1)}= 2 \pi {\rm i} \, 2 \, M(s_+) g^2 \phi(s_{\rm C}, m^2_-) \Delta_1(s_{\rm C}, m^2, s_+). 
  \label{eq:woolyit1}
  \ee
  Taking again the residue of both sides at the  second sheet pole of $M$, we obtain 
  \be 
  [\phi(s_{\rm C}, m^2_+) - \phi(s_{\rm C}, m^2_-)]^{(1)} = 2 \pi {\rm i} \, 2 \, g^2 \phi(s_{\rm C}, m^2_-) \Delta_1(s_{\rm C}, m^2, s_{\rm C}). 
  \label{eq:woolyit2}
  \ee 
  Referring to (\ref{eq:psi}), this can be written as 
  \be 
   [\phi(s_{\rm C}, m^2_+) - \phi(s_{\rm C}, m^2_-)]^{(1)} = 2 \pi {\rm i} \, 2 \, g^2 \phi(s_{\rm C}, m^2_-) \psi^{(1)}(s_{\rm C}, m^2, s_{\rm C}) 
   \label{eq:woolyit3}
   \ee
  where $\psi^{(1)}$ is the first iteration of the reduced $3 \to 3$ amplitude given by (\ref{eq:psiPsi}). Iterating the integral equation for our discontinuity then gives 
   \be 
   [\phi(s_{\rm C}, m^2_+) - \phi(s_{\rm C}, m^2_-)]= 2 \pi {\rm i} \, 2 \, g^2 \phi(s_{\rm C}, m^2_-) \psi(s_{\rm C}, m^2_+, s_{\rm C}). \label{eq:woolydiscend}
   \ee 
   
   Now we know that $\Delta_1/\sigma$ is proportional to the $s$-wave projection of the one-particle exchange diagram in the effective $3 \to 3$ process ``$\lambda^2 + 1 \to 1 + s$'' of figure 38. So allowing for this kinematical factor, we define the reduced particle + resonance  amplitude as 
   \be 
   R(s_{\rm C}, m^2_+, s_{\rm C}) = g^2 \frac{\psi(s_{\rm C}, m^2_+, s_{\rm C})}{\sigma(s_{\rm C}, m^2_+)}. \label{eq:defredR} 
   \ee 
   So finally our discontinuity across the particle + resonance branch cut is 
   \be 
   [\phi(s_{\rm C}, m^2_+) - \phi(s_{\rm C}, m^2_-)] = 2 \pi {\rm i} \, 2 \, \phi(s_{\rm C}, m^2_-) \sigma(s_{\rm C}, m^2) R(s_{\rm C}, m^2_+, s_{\rm C}), 
   \label{eq:woolydiscR}
   \ee 
   which we can represent as in figure 45. Note that
   \begin{figure} 
   \begin{center} 
  \includegraphics{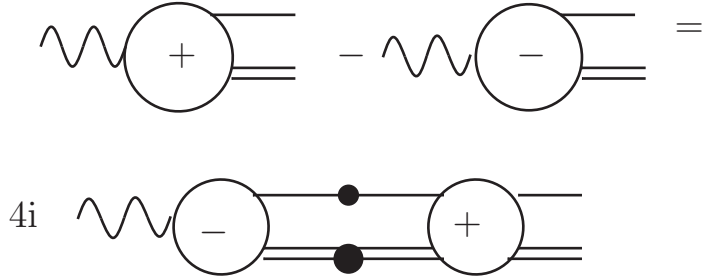} 
   \end{center} 
   \caption{The discontinuity across the woolly cut.} 
   \end{figure}  
   \be 
   \sigma(s_{\rm C}, m^2) = \{ [m^2 - (\sqrt{s_{\rm C}} +1)^2][m^2 - (\sqrt{s_{\rm C} } - 1)^2]\}^{1/2} /m^2, 
   \label{eq:sigcplx}
   \ee 
   which has the expected branch point at $(\sqrt{s_{\rm C}} +1)^2$, and is proportional to the phase space factor for a two-particle state with masses $\sqrt{s_{\rm C}}$ and 1. 
   
   The extra ``2'' in (\ref{eq:woolydiscR}) arises from the identical $t$ and $u$ channels in this toy model. In the case of just a single strong resonance + particle state, (\ref{eq:woolydiscR}) would take exactly the form of (\ref{eq:discF}), with $\rho$ replaced by $\sigma$. Thus our reduced amplitude $\phi(s_{\rm C}, m^2)$ (and of course the similar function $\Phi$) satisfies the expected quasi two-body discontinuity formula, in which the quasi two-body $\to$ quasi two-body scattering amplitude $R$ is self-consistently contained in the model, via the repeated rescatterings with one-particle exchange. The original isobar production amplitude $C(m^2)$, on the other hand, has been assumed to have no singularities in $m^2$. 
   
   Similar manipulations show that $R(s_{\rm C}, m^2, s_{\rm C})$ also has the branch point at $m^2=(\sqrt{s_{\rm C}} +1)^2$, with discontinuity 
   \be{\rm disc}_{m^2=(\sqrt{s_{\rm C}}+1)^2}\, R(s_{\rm C}, m^2, s_{\rm C}) = 4 \pi {\rm i} R(s_{\rm C}, m^2_+, s_{\rm C}) \sigma(s_{\rm C}, m^2) R(s_{\rm C}, m^2_-, s_{\rm C}), \label{eq:woollydiscR}
   \ee 
   which is the simple (quasi) 2-body discontinuity relation with $\rho$ replaced by $\sigma$. We can represent (\ref{eq:woollydiscR}) by figure 46. 
   \begin{figure}
   \begin{center}
   \includegraphics{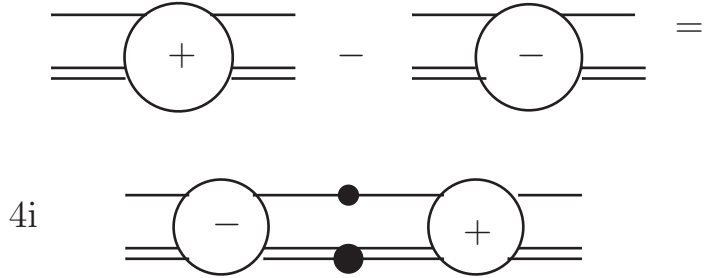}
   \end{center} 
   \caption{Discontinuity across the woolly cut of the quasi 2$\to$2 particle-resonance  amplitude $R$.}
   \end{figure}

   We may  regard this aspect of 3-body unitarity as the ``long-range'' part, associated as it is with repeated one-particle exchange between resonances. Viewing this dynamically, we might wonder whether such exchanges could produce a strong effect in the 3-body channel, perhaps even a  resonance. This is the same, of course, as asking whether the $m^2$-dependence of the amplitude $\Phi(s, m^2)$ could ever resemble a resonance. In the case of the truncated SVR (i.e. omitting the $\lambda^2 \leq 0$ part of the integrals), calculations showed that this is possible (Aitchison and Golding \cite{me78}, Pasquier \cite{P}, Parker \cite{ken}), though not for the physically relevant 
   $\pi-\pi$ amplitudes employed in these $3 \pi$ systems. But in any case, as noted earlier, such physical $ 3 \pi$ resonant states are generated by $q \bar{q}$ dynamics, not $ 3 \pi$ dynamics. 
   All the same, the extraction of the characteristics (e.g. pole positions and residues) of such resonances could be  substantially affected by $m^2$-dependent rescattering effects. 
   
   It seems natural to place the short-range dynamics leading to resonant states in the $m^2$ channel
   in the production function $C(m^2)$. Such states, once formed, decay to quasi two-body particle + resonance states, and we may parametrise such decays by a version of the $K$-matrix formalism of section 3, using the ``woolly'' phase space function $\sigma$. We would then end up (in this toy model) with an amplitude of the form 
   \be 
   (1-{\rm i} K \sigma)^{-1} P \, [ \Phi(s, m^2) + \Phi(t, m^2) + \Phi(u, m^2)], 
   \label{eq:Ffinal} 
   \ee 
   with $\sigma$ given by (\ref{eq:sigcplx}). Equation (\ref{eq:Ffinal}) represents the formation of an ``intrinsic'' state described by  $K$-matrix and $P$-vector poles, decaying as in the isobar model but with all pair-wise final state rescatterings included.
   
   \section{Conclusion}
   
   The main focus of these lectures has been on corrections to the isobar model for three-hadron final states, so as to incorporate the constraints of two-body unitarity and analyticity, primarily, and crossing symmetry somewhat indirectly. Initially the formalism appeared to be restricted to 
   imposing these constraints on the final state two-body channels only. This is in contrast to alternative approaches which treat the problem as an essentially three-body one, ensuring three-body unitarity as a primary object by the use of model Hamiltonians (for example \cite{Kam}) or via various relativistic scattering theory frameworks. 
   
   The ``two-body'' approach was originally introduced by Khuri and Treiman (K-T) \cite{KT}, in the context of two-particle final state interactions in the decay ${\rm K} \to \pi \pi \pi$, with emphasis on the two-body spectra. It is  well adapted to delivering the functions which correct the isobar model so as to satisfy the constraints of two-body unitarity and analyticity, only requiring as input the two-body amplitudes (on-shell, but extended as usual in dispersion theory beyond the physical region). There is no need for the complicated formalism of high spin fields in an effective Lagrangian, since the usual angular momentum decomposition of the isobar model is employed. The kinematics is relativistic throughout; admittedly, in the case of fermions it causes algebraic complications, but these will arise in any approach.  It seems fair to say that the constraints of two-body unitarity plus analyticity are ``minimal'', and least model-dependent, in the sense that they should be respected by any model. And they turn out to yield a surprising amount of structure. 
   
   Various versions of the K-T approach are available. The one I have  favoured here is what I call the single variable representation (SVR), where the amplitudes obey an integral equation involving only one integration variable. This form may be more convenient for calculations, but its real merit is that it is well suited to analysing the model's properties in the three-body channel. Somewhat surprisingly, this simple 
   ``two-body unitarity + analyticity'' model turns out to contain within it a three-body amplitude (consisting of pairwise rescatterings), in terms of which three-body unitarity is {\em also} satisfied. As originally suggested by Bonnevay \cite{georges1} \cite{georges2}, this can be traced to a form of crossing symmetry, in which the three-body decay amplitude ``$m  \to 1+ 2 + 3$'' is the crossed version (i.e. the analytic continuation in $m^2$) of  the two-body  amplitude ``$m +{\bar{1}} \to 2+3$''. The crucial three-body element in the SVR is the kernel function $\Delta_1$ (and similar objects in the non-zero spin cases), which is essentially the partial wave projection of the real particle exchange (RPE) process in the $3 \to 3$ sector - the quintessential feature of three-body dynamics. Arguments were given for retaining only the $\Delta_1$-type kernels in a truncated form of the SVR, represented diagrammatically by figure 39. 
   
   The KT approach is very simple in concept, and delivers isobar model correction factors in the form of numerical solutions depending on one subenergy variable (and the three-body mass), given only the two-body amplitudes. This technology, developed some 40-50 years ago, can be incorporated into fits to three-hadron final state physics. At the same time, of course, other approaches must continue to be pursued. It will be very interesting to see how they compare, in  cases where direct comparisons are possible. Perhaps we may be able to arrive at a convincingly more accurate version of that traditional work-horse, the isobar model. \\
   \vspace{1in}
   
   {\bf {\large{Acknowledgements}}}\\
   
   I am grateful to Adam Szczepaniak, Geoffrey Fox and Tim Londergan for inviting me to lecture at the 2015 International Summer Workshop in Reaction Theory at IU, and for organizing such a stimulating meeting. I thank JoAnne Hewett, Michael Peskin and Stan Brodsky for welcoming me as a visitor to the SLAC Theory Group, where this work was supported by the U.S. Department of Energy, Contract DE-AC02-76SF00515. I dedicate it to the memory of the late Georges Bonnevay, who died in a mountain accident in 1963, and who was my first instructor on the nursery slopes of complex surfaces.

\end{document}